\documentclass[aps,prb,twocolumn,superscriptaddress]{revtex4-1}

\usepackage{physics}
\usepackage{bbold}
\usepackage{lgrind}        
\usepackage{natbib}    
\usepackage{color}         
\usepackage[pdftex]{graphicx}      
\usepackage{epsf}          
\usepackage{bm}            
\usepackage{amsmath}
\usepackage{thumbpdf}
\usepackage{esint}
\usepackage[belowskip=-18pt,aboveskip=0pt]{caption}
\usepackage{subcaption}
\usepackage[colorlinks=true]{hyperref} 
\usepackage{placeins}
\usepackage{multirow}
\usepackage{multicolumn}
\usepackage[nodisplayskipstretch]{setspace}
\begin{document}

\title{Characterizing and Overcoming Surface Paramagnetism in Magnetoelectric Antiferromagnets}


\author{Sophie F. Weber}
\affiliation{Materials Theory, ETH Z\"{u}rich, Wolfgang-Pauli-Strasse 27, 8093 Z\"{u}rich, Switzerland}
\author{Nicola A. Spaldin}
\affiliation{Materials Theory, ETH Z\"{u}rich, Wolfgang-Pauli-Strasse 27, 8093 Z\"{u}rich, Switzerland}

\date{\today}
\begin{abstract}
We use a combination of density functional theory and Monte Carlo calculations to calculate the surface magnetization in magnetoelectric $\mathrm{Cr_2O_3}$ at finite temperatures. Such antiferromagnets, lacking both inversion and time-reversal symmetries, are required by symmetry to posses an uncompensated magnetization density on particular surface terminations. Here, we first show that the uppermost layer of magnetic moments on the $(001)$ surface remain paramagnetic at the bulk N\'{e}el temperature, bringing the theoretical estimate of surface magnetization density in line with experiment. We demonstrate that the lower surface ordering temperature compared to bulk is a generic feature of surface magnetization when the termination reduces the effective Heisenberg coupling. We then propose two methods by which the surface magnetization in $\mathrm{Cr_2O_3}$ could be stabilised at higher temperatures. Specifically, we show that the effective coupling of surface magnetic ions can be drastically increased either by a different choice of surface Miller plane, or by $\mathrm{Fe}$ doping. Our findings provide an improved understanding of surface magnetization properties in AFMs.
\end{abstract}


\maketitle
\indent Magnetoelectric (ME) antiferromagnets (AFMs) acquire a net magnetization $\mathbf{M}$ in response to an applied electric field $\mathbf{E}$, and conversely, a net electric polarization $\mathbf{P}$ in response to an applied magnetic field $\mathbf{H}$\cite{Astrov1960}. For the linear ME effect to manifest, an AFM must lack both inversion and time-reversal symmetries. This symmetry requirement implies another intriguing property of ME AFMs; namely, that certain surfaces must have a finite magnetic dipole density\cite{Belashchenko2010}. Such surface magnetization in ME AFMs has promising device applications, since the ME effect allows the bulk domain to be readily switched using electric fields in a constant magnetic field\cite{Ye2022}, and the direction of surface magnetization, which couples to the bulk AFM order parameter, can be directly detected\cite{Hedrich2021}. Additionally, surface magnetization plays a role in exchange bias coupling, extensively exploited in magnetic sensors and storage devices to pin the magnetization orientation of a ferromagnet (FM) by an adjacent AFM\cite{Nogues1999,Stamps2000}.\\
\indent An important question about surface magnetism is its degree of disorder close to the bulk N\'{e}el temperature $\mathrm{T}_N^{\mathrm{bulk}}$. Indeed, in the case of $\mathrm{Cr_2O_3}$ (chromia), a prototypical ME AFM viewed as a promising spintronics candidate due to its high N\'{e}el temperature of $\sim300$ $\mathrm{K}$\cite{Echtenkamp2021,Schlitz2018,Muduli2021,Ye2022}, theoretical predictions assuming that the bulk AFM order persists at the surface greatly overestimate the size of the $(001)$ surface magnetization density measured using nitrogen vacancy magnetometry\cite{Appel2019, Wornle2021, Spaldin2021}. This discrepancy is resolved if the outermost surface $\mathrm{Cr}$ moments are disordered at the measurement temperature, at or just below $\mathrm{T}_N^{\mathrm{bulk}}$\cite{Spaldin2021}. In general, a better understanding of the temperature dependence of surface magnetization in AFMs would facilitate quantitative comparison between theory and experiment, and could inform design of related spintronics devices.\\
\indent In this letter, we use a combination of density functional theory (DFT) and Monte Carlo (MC) calculations to explore the temperature dependence of surface magnetism, taking $\mathrm{Cr_2O_3}$ as an example. We show that partial to full disorder is a generic property of surface magnetization around the bulk ordering temperature when surface magnetic moments have fewer or smaller magnetic interactions than the bulk. We then propose two promising options for stabilizing the surface magnetization of $\mathrm{Cr_2O_3}$ at the bulk N\'{e}el temperature, first by using a Miller plane with a magnetic coupling close to bulk, and secondly by adding a monolayer of $\mathrm{Fe}$ on the $(001)$ surface.\\
\indent We first restate two key concepts, discussed in detail elsewhere\cite{Stengel2011,Spaldin2021}. The first regards the construction of an electrostatically stable, nonpolar surface termination for a given Miller plane $(h,k,l)$\cite{Stengel2011}.  A stable surface must have no bound charge, since a finite $\sigma_{\mathrm{surf}}$ implies a diverging electrostatic potential\cite{Nakagawa2006}. $\sigma_{\mathrm{surf}}$ is determined by the component of bulk electric polarization $\mathbf{P}_{\mathrm{bulk}}$ perpendicular to the surface\cite{King-Smith1993b}: $\mathbf{P}_{\mathrm{bulk}}\cdot \hat{\mathbf{n}}=\sigma_{\mathrm{surf}}$, where
$\hat{\mathbf{n}}$ is the unit surface normal. The periodicity of a bulk crystal implies that $\mathbf{P}_{\mathrm{bulk}}$ is only defined modulo a ``polarization quantum" which corresponds to translating one electron by a lattice vector\cite{King-Smith1993}. However, selecting a specific surface termination dictates a particular basis choice for the bulk unit cell (that which periodically tiles the semi-infinite solid containing the surface of interest), and hence a single value of $\mathbf{P}_{\mathrm{bulk}}$. In this case, $\sigma_{\mathrm{surf}}$ is single-valued, and a stable surface plane has $\mathbf{P}_{\mathrm{bulk}}\cdot \hat{\mathbf{n}}=\sigma_{\mathrm{surf}}=0$.\\
\indent The second point relates to the connection between the bulk ME multipolization tensor and the surface magnetization\cite{Spaldin2021}. The multipolization tensor is defined formally as $\mathcal{M}_{ij} =1/V\int r_i\mathbf{\mu}_j(\mathbf{r})d^3\mathbf{r}$, where $r_i$ is the $i\mathrm{th}$ cartesian component of position, $\mu_j(\mathbf{r})$ is the $j\mathrm{th}$ component of magnetization density at position $\mathbf{r}$, and $V$ is the unit cell volume. $\mathcal{M}$ describes first-order asymmetry in $\mathbf{\mu}(\mathbf{r})$ beyond the magnetic dipole\cite{Spaldin2013}. For materials in which $\mu(\mathbf{r})$ is localized around magnetic ions, $\mathcal{M}_{ij}$ can be approximated by:
\begin{equation}
\mathcal{M}_{ij}=\frac{1}{V}\sum_{\alpha}r^{\alpha}_im^{\alpha}_j,
\label{eq:mult_localmom}
\end{equation}
where the sum is over magnetic ions in the unit cell, and $\mathbf{m}^{\alpha}$ is the local magnetic moment of atom $\alpha$.\\
\indent The requirements for $\mathcal{M}$ to have nonzero components, that is, broken inversion and time-reversal symmetries, are identical to those for a nonzero linear ME response. Since a surface normal $\hat{\mathbf{n}}$ and an electric field $\mathbf{E}$ are both polar vectors, introducing a surface reduces the symmetry in the same way as applying an electric field in the bulk; therefore, ME AFMs must have nonzero surface magnetization\cite{Spaldin2021,Belashchenko2010,He2010}. By analogy with the surface charge density $\sigma_{\mathrm{surf}}$ resulting from the bulk polarization $\mathbf{P}_{\mathrm{bulk}}$, the bulk multipolization tensor $\mathcal{M}$ gives rise to a surface magnetic dipole density $\mu_{\mathrm{surf}}$\cite{Spaldin2021}, with the component $\mathcal{M}_{ij}$ giving the $\hat{i}$-oriented magnetization density on a surface whose normal is parallel to $\hat{j}$. Like $\mathbf{P}_{\mathrm{bulk}}$, the components of $\mathcal{M}$ are defined modulo a multipolization ``increment", corresponding to moving a magnetic ion by one lattice vector. But again, once a specific Miller plane and atomic termination are selected, the origin of the bulk unit cell and thus the value of multipolization are fixed. Therefore, the surface magnetization associated with $\mathcal{M}$ is a single-valued quantity in the limit of bulk-like order of the surface magnetic moments (see supplement for further discussion on the connection between $\mathcal{M}$ and surface magnetization).\\
\begin{figure}
\includegraphics{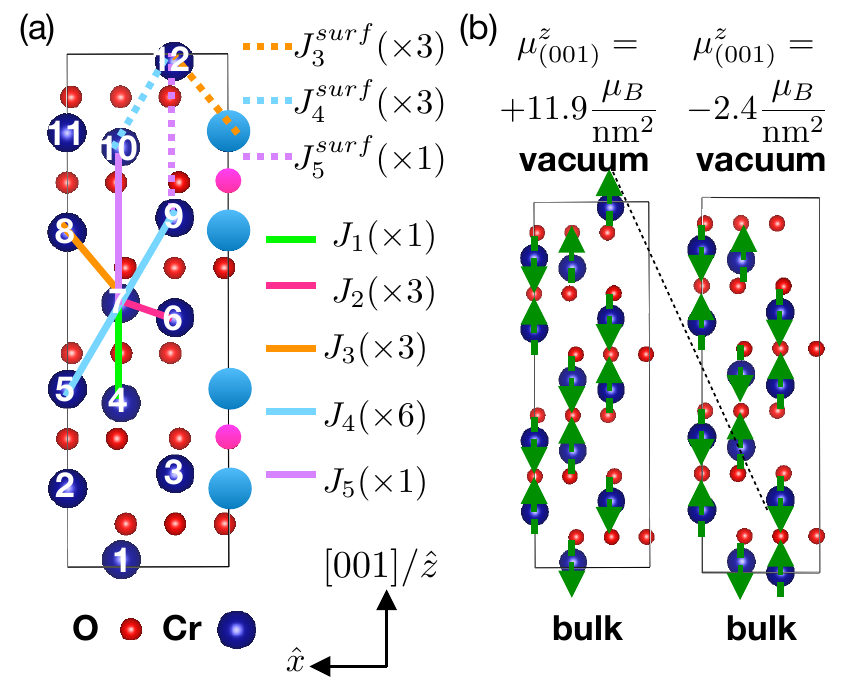}
\caption{\label{fig:Cr2O3_struct}(a) Hexagonal unit cell of $\mathrm{Cr_2O_3}$ with nearest neighbor Heisenberg couplings indicated for bulk (solid lines) and $(001)$ surface (dashed lines) $\mathrm{Cr}$s. Numbers in parentheses indicate the coupling degeneracies (not all nearest neighbors are visible in the diagram). Lighter colored atoms are in the adjacent unit cell. White numbers label the $12$ magnetic $\mathrm{Cr}$ in the $(001)$ slab for ease of discussion. (b) Left: unit cell which defines the nonpolar $(001)$ surface with ground state AFM magnetic ordering. Right: unit cell used to calculate surface magnetization if the top moment of the nonpolar surface is disordered.} 
\end{figure}
\begin{table*}
\caption{Heisenberg coupling constants, with degeneracies in parentheses, and total effective coupling $\lambda_J^i$ calculated in this work for bulk $\mathrm{Cr_2O_3}$, for a $\mathrm{Cr}$ ion on a $(001)$ surface (using the bulk values and those computed from a relaxed slab respectively), for a $\mathrm{Cr}$ on a $(\bar{1}02)$ surface, and $\mathrm{Cr}$-$\mathrm{Fe}$ couplings for $[001]$ $\mathrm{Cr_2O_3}$ with an $\mathrm{Fe}$ monolayer.}\label{tab:couplings}
	\begin{tabular}{| c | c | c | c | c | c | c | }
 \hline
& \multicolumn{1}{|c|}{bulk}
&
 \multicolumn{2}{|c|}{$(001)$ surface} 
 & 
 \multicolumn{2}{|c|}{$(\bar{1}02)$ surface} 
 &
 \multicolumn{1}{|c|}{$\mathrm{Fe}$ monolayer on $(001)$ surface}\\ \hline
& & bulk & relaxed & bulk & relaxed & relaxed\\
 \hline
$J_1$ ($\mathrm{meV}$) & -10.46 (1)&-&-& -10.46 (1) & -17.16 (1) &-\\
\hline
$J_2$ ($\mathrm{meV}$) & -7.88 (3)&-&-& -7.88 (2) & -9.42 (2) &-\\
\hline
$J_3$ ($\mathrm{meV}$) & +0.86 (3)& +0.86 (3)& -0.15 (3) & +0.86 (1) &+0.43 (1)&-30.81 (3) \\
\hline
$J_4$ ($\mathrm{meV}$) & +1.22 (6)&+1.22 (3) & +4.44 (3) & +1.22 (5) &+0.43$^a$ (4)/+3.40$^b$ (1)&-19.77 (3) \\
\hline
$J_5$ ($\mathrm{meV}$) & -1.41 (1)& -1.41 (1)& -0.39 (1) &-&-& -3.63 (1)\\
\hline
$\lambda_J$ ($\mathrm{meV}$) & 40.25 & 2.48& 14.18 & 31.47 & 40.16 &155.33\\
\hline
	\end{tabular}
\end{table*}
\indent \emph{Results and Discussion.}--$\mathrm{Cr_2O_3}$ crystallizes in the corundum structure with magnetic space group $\mathrm{R\bar{3}'c'}$ [161]\cite{Fechner2018}. Figure \ref{fig:Cr2O3_struct}(a) shows the $12$-$\mathrm{Cr}$ unit cell in the hexagonal setting. Bulk $\mathrm{Cr_2O_3}$ adapts an ``up down up down'' ordering of the $\mathrm{Cr}$ magnetic moments along $[001]$ as shown for the unit cells in Figure \ref{fig:Cr2O3_struct}(b). This ground state order is well described\cite{Samuelsent1970,Shi2009} by a Heisenberg Hamiltonian, 
\begin{equation}
\mathcal{H}_{\mathrm{Heis}}=-\sum\limits_{<i,j>}J_{i,j}(\mathbf{e}_i\cdot\mathbf{e}_j),
\label{eq:HeisH}
\end{equation}
that includes coupling up to the fifth nearest neighbors, where $\mathbf{e}_i$ is the unit vector parallel to the local magnetic moment of the $\mathrm{Cr}$ ion at site $i$, and $J_{i,j}$ is the Heisenberg coupling constant between spins $i$ and $j$. The couplings $J_1$-$J_5$, where $J_n$ denotes the coupling for the $\mathrm{n}^{th}$ nearest neighbor, are depicted in Figure \ref{fig:Cr2O3_struct}(a). The quantitative values of $J_1$-$J_5$ for bulk $\mathrm{Cr_2O_3}$, which we calculate with the method outlined in reference \citenum{Xiang2011} using first-principles DFT+U as implemented in the VASP software\cite{Kresse1996}, are given in Table \ref{tab:couplings}. Our values are in good agreement with previous DFT calculations using similar parameters\cite{Shi2009}. The magnetism is dominated by the strong AFM $J_1$ and $J_2$ couplings.\\
\indent  We first review magnetism on the $(001)$ surface of vacuum-terminated chromia. The bulk unit cell with a single terminating $\mathrm{Cr}$ on the left-hand side of Figure \ref{fig:Cr2O3_struct}(b) defines the nonpolar $(001)$ surface according to the formula $\mathbf{P}_{\mathrm{bulk}}=1/V\sum_iZ_i\mathbf{r}_i$, where $Z_i$ is the formal ionic charge ($+3$ and $-2$ for $\mathrm{Cr}$ and $\mathrm{O}$ respectively), and $\mathbf{r}_i$ the position of atom $i$ in the unit cell. If we assume all $\mathrm{Cr}$ magnetic moments are fully polarized along $[001]$ with the bulk AFM order, using the formal value $3\mathrm{\mu_B}$ for $\mathrm{Cr^{3+}}$ and the fractional coordinates (given in the supplement) in the hexagonal cell, equation \ref{eq:mult_localmom} yields a $+\hat{z}/[001]$-oriented $(001)$ surface magnetization of $11.9$ $\mathrm{\mu_B}/\mathrm{nm^2}$ for the magnetic domain depicted (all other components of the multipolization tensor are zero within the local moment approximation; small $(1,1)$ and $(2,2)$ components are symmetry-allowed if one uses the exact integral form\cite{Urru2022}). The energetically equivalent AFM domain in which the directions of all magnetic moments are reversed has a value of equal magnitude and opposite sign.\\
\indent As mentioned previously, this theoretical predication overestimates measurements of $(001)$ $\mathrm{Cr_2O_3}$ surface magnetism using scanning nitrogen vacancy magnetometry\cite{Appel2019,Wornle2021}, which yield values between $1.6$  to $2.3$ $\mathrm{\mu_B}/\mathrm{nm^2}$ (the sign of magnetization cannot be directly determined). Recall however that the $11.9$ $\mathrm{\mu_B}/\mathrm{nm^2}$ value is calculated assuming that all $\mathrm{Cr}$ magnetic moments are fully ordered along $[001]$. Looking at the outermost $\mathrm{Cr}$ for the nonpolar termination in Figure \ref{fig:Cr2O3_struct}(a), we see that it lacks $J_1$ and $J_2$  nearest neighbors, only retaining the smaller $J_3$-$J_5$ couplings. From a mean-field argument, the ordering temperature for a given magnetic moment at site $i$ is proportional to $\lambda^i_JS_i$\cite{Mostovoy2010}, where $S_i$ is the spin value and the total effective Heisenberg coupling for site $i$ is
\begin{equation}
\lambda^i_J=\sum\limits_j J_{ij}\times(\hat{e}_i\cdot \hat{e_j}).
\label{eq:Jsum}
\end{equation}
Using the $J$ values in Table \ref{tab:couplings} calculated for bulk $\mathrm{Cr_2O_3}$, $\lambda_J$ for a bulk $\mathrm{Cr}$ spin is $\lambda_J^{\mathrm{bulk}}=-J_1-3J_2-3J_3+6J_4-J_5=40.25\mathrm{meV}$, whereas the $\mathrm{Cr}$ on the $(001)$ surface ($\mathrm{Cr}$ $12$ with the convention in Figure \ref{fig:Cr2O3_struct}(a)) has $\lambda_J^{\mathrm{surf}}=-3J_3+3J_4-J_5= 2.48\mathrm{meV}$.  Thus, $T_N^{\mathrm{surf}}/T_N^{\mathrm{bulk}}=0.06$, implying that for the room temperature $T\sim293\mathrm{K}$ magnetometry measurements, just below $\mathrm{T_N^{bulk}}\sim 300 \mathrm{K}$, we expect the surface $\mathrm{Cr}$ to be paramagnetic. Taking into account this magnetic dead layer, a more appropriate basis for predicting the surface magnetization is that shown on the right-hand side of Figure \ref{fig:Cr2O3_struct}(b), corresponding to removing the surface magnetic moment by displacing it downwards one $\mathbf{c}$ lattice vector. Recalculating the $(3,3)$ component of $\mathcal{M}$ using the $\mathrm{Cr}$ positions of this new unit cell yields $\mu_{(001)}^z=-2.4$ $\mathrm{\mu_B}/\mathrm{nm^2}$, in good agreement with experiment.\\
\begin{figure}
\includegraphics{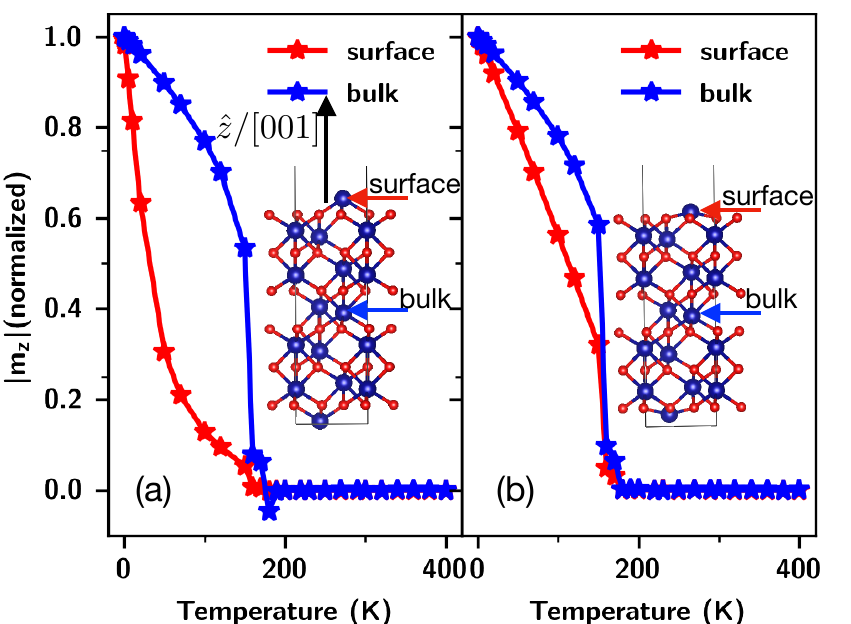}
\caption{\label{fig:MT_001surf}(a) Normalized magnetization along $\hat{z}/\mathrm{[001]}$ as a function of temperature for a bulk $\mathrm{Cr}$ in the center of a $[001]$-oriented slab, and for the  surface $\mathrm{Cr}$ (labeled on the structure in the inset). Here the Heisenberg couplings calculated from bulk $\mathrm{Cr_2O_3}$ (first column of Table \ref{tab:couplings}) are used for both surface and bulk $\mathrm{Cr}$ moments. (b) $m_z$ using couplings calculated from a relaxed slab for the surface $\mathrm{Cr}$.} 
\end{figure}
\indent To confirm our analysis, we examine the temperature dependence of magnetization by performing Monte Carlo (MC) simulations as implemented in the UppASD spin dynamics package\cite{Skubic2008} of a $[001]$-oriented $\mathrm{Cr_2O_3}$ slab using a $42\times42\times1$ supercell of the $12$-$\mathrm{Cr}$-atom hexagonal unit cell, having checked that this thickness, with six $\mathrm{Cr_2O_3}$ layers, is sufficient to capture both bulk and surface behavior. We enforce in-plane periodic boundary conditions and vacuum boundary conditions along $[001]$. Figure \ref{fig:MT_001surf}(a) shows the absolute value of the $\hat{z}$ component of bulk magnetization as a function of temperature, calculated by averaging the projected $m_z$ of the $6^{\mathrm{th}}$ $\mathrm{Cr}$ sublattice in the center of our unit-cell thick slab, compared to the averaged $m_z$ of the terminating $\mathrm{Cr}$s on the nonpolar $(001)$ surface.  We also confirmed that all sublattices other than the outermost $\mathrm{Cr}$ on both sides of the slab have bulk-like $m_z(T)$ behavior. We see that, whereas the center ``bulk" $\mathrm{Cr}$ exhibits the normal Langevin-like $m_z(T)$ curve, the surface magnetization falls off rapidly with increasing temperature and is negligible at $T_N^{\mathrm{bulk}}$, consistent with earlier combined DFT-MC calculations\cite{Wysocki2012}. Note that our calculated Heisenberg constants lead to a significant underestimate of $T_N^{\mathrm{bulk}}$ ($T_N^{\mathrm{bulk}}\sim170$ $\mathrm{K}$ based on Figure \ref{fig:MT_001surf}); this has been observed in previous DFT-MC calculations of $\mathrm{Cr_2O_3}$\cite{Kota2013}. \\
\indent While $m_z(T)$ in Figure \ref{fig:MT_001surf}(a) for both surface and bulk $\mathrm{Cr}$s are computed using the DFT $J$ values calculated with bulk $\mathrm{Cr_2O_3}$, atomic relaxation can lead to significant renormalization of the surface couplings. The third column of Table \ref{tab:couplings} shows the values of $J_3$, $J_4$ and $J_5$ for the surface $\mathrm{Cr}$ computed using a $[001]$ vacuum-terminated $12$-$\mathrm{Cr}$-thick slab which we structurally relax within DFT. The effective coupling for the surface $\mathrm{Cr}$ when taking relaxation into account is  $\lambda_J^{\mathrm{surf,relaxed}}=14.2\mathrm{meV}$. Figure \ref{fig:MT_001surf}(b) shows $m_z(T)$ for the surface $\mathrm{Cr}$ with these relaxed values (we keep the remaining $J$s for the the ten non-surface $\mathrm{Cr}$ set to bulk values, having checked that the coupling renormalization for these ions upon relaxation negligibly affects the results). 
While the surface $m_z$ is still disordered at $T_N^{\mathrm{bulk}}$, the increased $\lambda_J^{\mathrm{surf}}$ leads to a roughly linear decrease of $m_z$ with increasing $T$, as opposed to the exponential-like falloff in Figure \ref{fig:MT_001surf}(a).\\
 \begin{figure}
\includegraphics{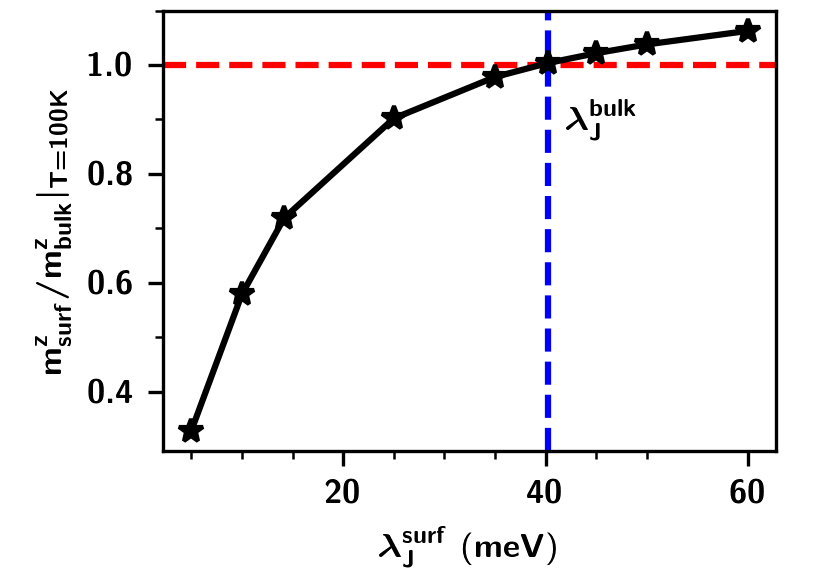}
\caption{\label{fig:J_eff} Ratio of $m_z^{\mathrm{surf}}$ to $m_z^{\mathrm{bulk}}$ at $T=100$ $\mathrm{K}$ ($\sim T_N^{\mathrm{bulk}}/2$) as a function of $\lambda_J^{\mathrm{surf}}$. $\lambda_J^{\mathrm{surf}}$ is fixed in the MC simulations by setting the three $J_4^{\mathrm{surf}}$ to $\lambda_J^{\mathrm{surf}}/3$ and all other surface couplings to zero. The dashed blue line shows the value of $\lambda_J^{\mathrm{bulk}}$.}
\end{figure}
 \indent To determine the dependence of the surface magnetism on $\lambda_J^{\mathrm{surf}}$ in detail, we next vary $\lambda_J^{\mathrm{surf}}$ manually in the MC simulations by fixing each of the three $J_4^{\mathrm{surf}}$ to one third the desired $\lambda_J^{\mathrm{surf}}$, while setting all other surface couplings to zero. For each value of $\lambda_J^{\mathrm{surf}}$ we calculate $\frac{m_z^{\mathrm{surf}}}{m_z^{\mathrm{bulk}}}$ at $T=100$ $\mathrm{K}$; the result is plotted in Figure \ref{fig:J_eff}. We choose $100$ $\mathrm{K}$ as a representative temperature because it is roughly $T_N^{\mathrm{bulk}}/2$ ($\lambda_J^{\mathrm{surf}}$ marginally affects $T_N^{\mathrm{bulk}}$ due to the finite slab size, thus $T_N^{\mathrm{bulk}}/2$ ranges from $85$ to $110$ $\mathrm{K}$ for the range of $\lambda_J^{\mathrm{surf}}$ in Figure \ref{fig:J_eff}). $\frac{m_z^{\mathrm{surf}}}{m_z^{\mathrm{bulk}}}|_{T=100 \mathrm{K}}$ increases monotonically with $\lambda_J^{\mathrm{surf}}$ and matches the bulk magnetization ($\frac{m_z^{\mathrm{surf}}}{m_z^{\mathrm{bulk}}}=1$) roughly when $\lambda_J^{\mathrm{surf}}$ equals $\lambda_J$ for the bulk $\mathrm{Crs}$ (dashed blue line). Therefore, engineering $\lambda_J^{\mathrm{surf}}$ to be close to $\lambda_J^{\mathrm{bulk}}$ can be taken as a criterion for obtaining bulk-like temperature dependence of surface magnetization. Moreover, if the Heisenberg $J$s for a material are known, one can quickly calculate $\lambda_J^{\mathrm{surf}}$ and estimate how much $m^{\mathrm{surf}}$ is likely to be reduced relative to bulk magnetization.\\
 \begin{figure}
\includegraphics{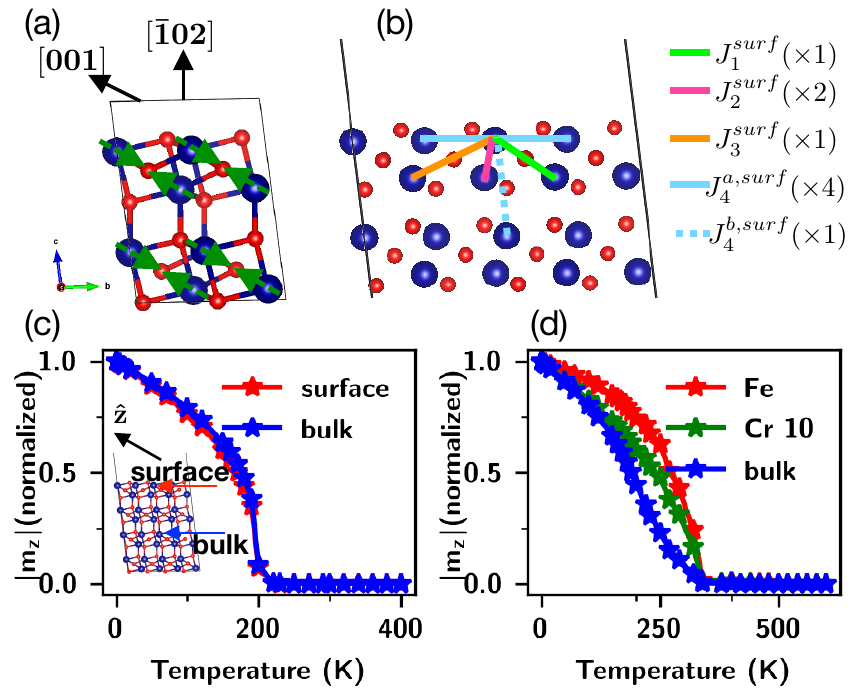}
\caption{\label{fig:stabilize}(a) Bulk unit cell defining nonpolar $(\bar{1}02)$ surface of $\mathrm{Cr_2O_3}$, with $[001]$ oriented bulk magnetic order. (b) Heisenberg couplings for $\mathrm{Cr}$ ions on the $(\bar{1}02)$ surface. (c) $|m_z(T)|$ (with $z$ along the bulk $[001]$ easy axis) for $\mathrm{Cr}$ in bulk and on a $(\bar{1}02)$ surface (using relaxed surface couplings). (d) $|m_z(T)|$ for $\mathrm{Fe}$, $\mathrm{Cr}$ $10$, and a central bulk $\mathrm{Cr}$ (positions $12$, $10$ and $6$ respectively from Figure \ref{fig:Cr2O3_struct}(a)).}
\end{figure}
 \indent We now discuss two approaches, which can also be applied to other ME AFMs, for stabilizing surface magnetization in $\mathrm{Cr_2O_3}$ at higher temperatures. The first is to use a surface corresponding to a different Miller plane for which $\lambda_J^{\mathrm{surf}}$ is close to $\lambda_J^{\mathrm{bulk}}$. We demonstrate this for $\mathrm{Cr_2O_3}$ in Figure \ref{fig:stabilize}. Figure \ref{fig:stabilize}(a) shows the unit cell corresponding to the non-polar termination of a $(\bar{1}02)$ surface, which has a large $\lambda_J^{\mathrm{surf}}$ and a non-negligible theoretical surface magnetization density. Specifically, for the domain shown, Equation \ref{eq:mult_localmom} predicts an out-of-plane (in-plane) magnetization component of $-4.75$ $\mathrm{\mu_B}/\mathrm{nm^2}$ ($+7.55$ $\mathrm{\mu_B}/\mathrm{nm^2}$) respectively on the $(\bar{1}02)$ surface (see supplement).\\
 \indent The couplings and degeneracies, shown in Figure \ref{fig:stabilize}(b), retained by the outermost $\mathrm{Cr}$ and corresponding $\lambda_J^{\mathrm{surf}}$ for the $(\bar{1}02)$ surface are given in Table \ref{tab:couplings} both with DFT values calculated from bulk $\mathrm{Cr_2O_3}$  and with surface couplings calculated from a relaxed $[\bar{1}02]$-oriented slab. The $a$ and $b$ superscripts refer to the two $J_4$ couplings depicted in Figure \ref{fig:stabilize}(b) which become inequivalent upon relaxation (we double the unit cell in the surface plane in order to show all couplings). The overall $\lambda_J^{\mathrm{surf}}$ is $31.47$ ($40.16$) $\mathrm{meV}$ for bulk (relaxed) coupling values. Even using the bulk values for $(\bar{1}02)$ surface moments, the surface magnetization is nearly bulk-like, and with the relaxed values leading to $\lambda_J^{\mathrm{surf}}\sim\lambda_J^{\mathrm{bulk}}$, bulk and $(\bar{1}02)$ surface $m_z(T)$ lie on top of each other (Figure \ref{fig:stabilize}(c)).\\
 \indent Extensive research has been devoted to the application of $[001]$ $\mathrm{Cr_2O_3}$ films in spintronic memory devices, where the AFM bulk domain serves as a logical bit whose direction can be read out by the sign of the surface magnetization (this is usually determined indirectly via the sign of the hysteresis loop shift, i.e. exchange bias, in an adjacent FM\cite{Borisov2005,He2010,Ye2022}). Our results imply that magnetism on the $(\bar{1}02)$ surface in chromia is strongly coupled to the underlying bulk AFM domain, even at $T_N^{\mathrm{bulk}}$, in contrast to the $(001)$ surface where the surface is essentially paramagnetic at room temperature. Thus, a $\mathrm{Cr_2O_3}$-based device with a $(\bar{1}02)$ rather than $(001)$ surface plane might be a more robust option for memory applications. More fundamentally, a comparison of exchange bias properties for the $(001)$ and $(\bar{1}02)$ surfaces could shed light on the underlying mechanism.\\
 \indent Our second proposed method for stabilizing surface magnetization involves chemical substitution. We take the $\mathrm{Cr_2O_3}$ $(001)$ surface, and deposit a monolayer of $\mathrm{Fe}$ on top (substituting $\mathrm{Cr}$ in the $12^{\mathrm{th}}$ position in Figure \ref{fig:Cr2O3_struct}(a)). Since $\mathrm{Fe}$ adopts a $3+$ valence state, this structure is nonpolar and stable (see supplement for further discussion). Crucially, while the $\mathrm{Cr}$-$\mathrm{Cr}$ $J_3$-$J_5$ are negligible compared to $J_1$ and $J_2$, prior DFT studies using a $\mathrm{Cr_2O_3}$-$\mathrm{Fe_2O_3}$ heterostructure indicate that the $J_3$ and $J_4$ $\mathrm{Cr}$-$\mathrm{Fe}$ couplings at the interface are tens of $\mathrm{meV}$\cite{Kota2014}. The difference in strengths and signs of $\mathrm{Cr}$-$\mathrm{Cr}$ and $\mathrm{Cr}$-$\mathrm{Fe}$ couplings in the corundum structure can be attributed to the relative $e_g$-$t_{2g}$ occupation of the $\mathrm{Cr^{3+}}$ $(t_{2g}^3,e_g^0)$ and $\mathrm{Fe^{3+}}$ $(t_{2g}^3,e_g^2)$ ions, combined with the coupling angles via oxygen\cite{Kota2013} In the final column of Table \ref{tab:couplings}, we show our results for the surface $\mathrm{Fe}$-$\mathrm{Cr}$ couplings calculated using a relaxed $\mathrm{Cr_2O_3}$ slab terminated on one side with $\mathrm{Fe}$. $J_3^{\mathrm{surf,Cr-Fe}}$ and $J_4^{\mathrm{surf,{}Cr-Fe}}$ are even larger than the $J_1^{\mathrm{Cr-Cr}}$ and $J_2^{\mathrm{Cr-Cr}}$ that are dominant in $\mathrm{Cr_2O_3}$ bulk.\\
 \indent Figure \ref{fig:stabilize}(d) shows the absolute value of $m_z(T)$ for the surface $\mathrm{Fe}$ monolayer and center $\mathrm{Cr}$ bulk, as well as for ``$\mathrm{Cr}$ $10$" (according to the labeling in Figure \ref{fig:Cr2O3_struct}(a)) which is coupled to the $\mathrm{Fe}$ via $J_4^{\mathrm{surf,{} Fe-Cr}}$ (note that $\mathrm{Cr}$ $10$ reverses its orientation from that in bulk due to the strong AFM $J_4^{\mathrm{surf,{}Fe-Cr}}$ coupling). The $\mathrm{Cr}$s coupled directly to the $\mathrm{Fe}$ monolayer have magnetization intermediate between those of the deeper bulk $\mathrm{Cr}$ and of $\mathrm{Fe}$.\\
 \indent A notable feature of the surface $\mathrm{Fe}$ monolayer magnetization in Figure \ref{fig:stabilize}(d) is that $m_z^{\mathrm{Fe,surf}}$ (and the $\mathrm{Cr}$ directly below) order at higher temperatures than the bulk $\mathrm{Cr}$, making it an attractive test case for fundamental research in paramagnetic bulk materials with surface magnetic order\cite{Dedkov2007, Rosenberg2012, Schulz2019}. Moreover, if scanning nitrogen vacancy magnetometry measurements of $\mathrm{Fe}$-capped $\mathrm{Cr_2O_3}$ could be compared at temperatures just above (where only $\mathrm{Fe}$ and the top-most $\mathrm{Cr}$ are ordered) and below $T_N^{\mathrm{bulk}}$, monitoring how the measured surface magnetization changes would provide a clear indication of the technique's depth resolution.\\
\indent In summary, we have examined finite temperature properties of surface magnetization in AFMs using ME $\mathrm{Cr_2O_3}$ as an example. Our combined DFT-MC calculations demonstrate that disorder of surface magnetic ions at $T_N^{\mathrm{bulk}}$ likely explains the discrepancy between theoretical and experimental surface magnetization estimates on $(001)$ $\mathrm{Cr_2O_3}$. We establish a framework for assessing the relative ordering temperature of surface and bulk magnetization based on effective Heisenberg couplings. Finally, we have discussed two options for stabilizing surface magnetism, which would allow for higher temperature operation of relevant spintronic devices. We hope this work stimulates efforts, both theoretical and experimental, to better understand and characterize surface magnetization in AFMs.\\

\begin{acknowledgments}
We thank Xanthe Verbeek, Tara To\v{s}i\'{c}, Kai Wagner, Paul Lehman, Patrick Maletinsky, Sayantika Bhowal and Andrea Urru for useful discussions. This work was funded by the ERC under the European Union's Horizon 2020 research and innovation program project HERO with grant number 810451. Calculations were performed at the Swiss National Supercomputing Centre (CSCS) under project number s889 and on the EULER cluster of ETH Z\"{u}rich.
\end{acknowledgments}

\begin{appendix}
\section{Density Functional Theory calculation details}
In order to calculate the relaxed structures and Heisenberg coupling constants for $\mathrm{Cr_2O_3}$ in this work, we use density functional theory (DFT), employing the Vienna \emph{ab initio} simulation package (VASP)\cite{Kresse1996} with the localized density approximation (LDA) within the projector augmented wave method (PAW)\cite{Blochl1994a}. We use the standard VASP PAW pseudopotentials with the following valence electron configurations: $\mathrm{Cr}$ $(3p^64s^23d^5)$, $\mathrm{O}$ $(2s^23p^4)$, and $\mathrm{Fe}$ $(4s^23d^6)$ (for calculations of $\mathrm{Fe}$-capped $\mathrm{Cr_2O_3}$). We use collinear spin-polarized calculations, neglecting spin-orbit coupling except when calculating the magnetocrystalline anisotropy. We use an energy cutoff of $800$ $\mathrm{eV}$ for our plane wave basis set, and a Gamma-centered $9 \times9\times5$ k-mesh for the bulk $30$-atom hexagonal unit cell of $\mathrm{Cr_2O_3}$. To model the $(001)$ ($(\bar{1}02)$) $\mathrm{Cr_2O_3}$ surfaces we use hexagonal (monoclinic) cells with $9\times9\times1$ ($11\times6\times1$) Gamma-centered k-meshes with $15$ $\mathrm{\AA}$ vacuum in the direction of the surface normals. We use the tetrahedron method\cite{Blochl1994a} for Brillouin zone integrations. We find that these parameters lead to total energy convergence of $<1$ $\mathrm{meV}$ per formula unit. We relax all structures, both bulk and slabs, until forces on all atoms are less than $0.01$ $\mathrm{eV}/\mathrm{\AA}$.\\
\indent To approximately capture the localized nature of $3d$ electrons in $\mathrm{Cr}$, we add a Hubbard U correction\cite{Anisimov1997} using the rotationally invariant method by Dudarev et al.\cite{Dudarev1998}. We set $\mathrm{U}=4$ $\mathrm{eV}$ based on prior DFT+U work on $\mathrm{Cr_2O_3}$ with $\mathrm{U}=4$, $\mathrm{J}\sim0.5$ $\mathrm{eV}$\cite{Shi2009,Fechner2018} (we also use $\mathrm{U}=4$ on the $\mathrm{Fe}$ $\mathrm{d}$ states for calculations of the $\mathrm{Fe}$ capped $[001]$ chromia slab). We find that including the Hund's coupling $\mathrm{J}=0.5$ $\mathrm{eV}$ does not significantly affect the computed Heisenberg coupling constants, thus we only use the Hubbard $\mathrm{U}$. As mentioned in the main text, with this $\mathrm{U}$ value our calculated couplings lead to an underestimated bulk N\'{e}el temperature $T_N^{\mathrm{bulk}}$ from the Monte Carlo simulations, which was also the case for a prior DFT-MC study of chromia using a similar U value\cite{Kota2014}. Using a smaller $\mathrm{U}$ ($\mathrm{U}=2$ $\mathrm{eV}$ for reference \citenum{Mostovoy2010}) would lead to larger Heisenberg couplings (due to decreased localisation), and hence a $T_N^{\mathrm{bulk}}$ closer to experiment\cite{Mostovoy2010}. However, we choose $\mathrm{U}=4$ $\mathrm{eV}$ because with this value we achieve the correct sign of magnetocystalline anisotropy energy (MAE), i.e. easy axis along the hexagonal $[001]$ direction\cite{Dudko1971,Tobia2010}. With $\mathrm{U}=2$ $\mathrm{eV}$ on the other hand, we calculate a qualitatively incorrect easy plane. Thus, we believe the higher $\mathrm{U}$ value overall better describes the magnetic properties of $\mathrm{Cr_2O_3}$. We also note that the \emph{relative} values of $J_1$ through $J_5$ are similar for a wide range of $\mathrm{U}$ as can be seen from reference \citenum{Shi2009}. Thus, the primarily qualitative conclusions drawn in our work hold in spite of the $T_N^{\mathrm{bulk}}$ underestimate.\\
\indent We compute the Heisenberg couplings for both bulk and slab structures using the method outlined in reference \citenum{Xiang2011}. Essentially, the coupling between two specific sites $i$ and $j$ is calculated from four total energy calculations in which the magnetic moments on these two sites are set to $(i,j)=(\uparrow\uparrow), (\uparrow\downarrow),(\downarrow\uparrow),(\downarrow\downarrow)$, with moments on all other sites in the unit cell kept constant. In the case of the relaxed vacuum-terminated structures, this method allows us to calculate the Heisenberg couplings for each $\mathrm{Cr}$ site and thus differentiate between the values for the surface $\mathrm{Cr}$ and for  $\mathrm{Cr}$ in the center of the slab which retain couplings close to the results from bulk. 
\section{Monte Carlo calculation details}
To explore the temperature dependence of surface magnetism in $\mathrm{Cr_2O_3}$, we use Monte Carlo (MC) simulations as implemented in the UppASD\cite{Skubic2008} spin dynamics package. We use supercells with $42\times42\times1$ ($38\times19\times1$) magnetic unit cells for the $(001)$ ($(\bar{1}02)$) surfaces, for a total of $21168$ ($23104$) magnetic atoms in the simulation box. We use periodic boundary conditions in the in-plane $\mathbf{a}$ and $\mathbf{b}$ directions  and vacuum boundary conditions in the out-of-plane $\mathbf{c}$ direction. To test convergence of our results, we also performed MC simulations with a simulation box doubled along the $\mathrm[001]$ surface normal, i.e. $24$-$\mathrm{Cr}$ tall as opposed to the $12$-$\mathrm{Cr}$ tall box used in the main manuscript. We found that the projected $m_z(T)$ for both surface $\mathrm{Cr}$ and $\mathrm{Cr}$ sublattices in the center of the slab did not change noticeably upon doubling the height; thus, the $12$-$\mathrm{Cr}$ tall box ($16$-$\mathrm{Cr}$ tall for the $(\bar{1}02)$ surface) is sufficiently thick to capture behavior of both bulk and surface magnetization. To prevent the magnetization axis from drifting, we add a uniaxial magnetoanisotropy energy (MAE) of $0.11$ $\mathrm{meV}$ per $\mathrm{Cr}$ along the $\hat{z}$/$[001]$ direction, along the lines of previous studies\cite{Mostovoy2010}. The experimental MAE of bulk $\mathrm{Cr_2O_3}$, as well as the value we calculate with DFT+U (about $4$ $\mu\mathrm{eV}$ per $\mathrm{Cr}$) is two orders of magnitude smaller\cite{Dudko1971,Tobia2010}. However, due to the finite size of the simulation box in the MC simulations the MAE must be scaled up to prevent unphysical fluctuations of the magnetization axis. Simulations at each temperature were performed with $2\times 10^5$ initial steps to bring the system to thermal equilibrium, and $N_{MC}=2\times 10^5$ subsequent MC iterations during which system properties were evaluated. The $\hat{z}$ component of magnetization for a given sublattice $i$ (where the number of sublattices is simply the number of magnetic ions in the magnetic unit cell) is calculated as 
\begin{equation}
m_z^i=\frac{1}{N_{cells}}\sum_j^{N_{cells}}m_{z,j}^i,
\label{eq:mz}
\end{equation}
where $N_{cells}$ is the number of magnetic cells in the MC simulation box ($42\times42=1764$ and $32\times19=722$ for the two surfaces studied).
\section{Multipolization tensors for different surfaces}
\begin{figure*}
\includegraphics{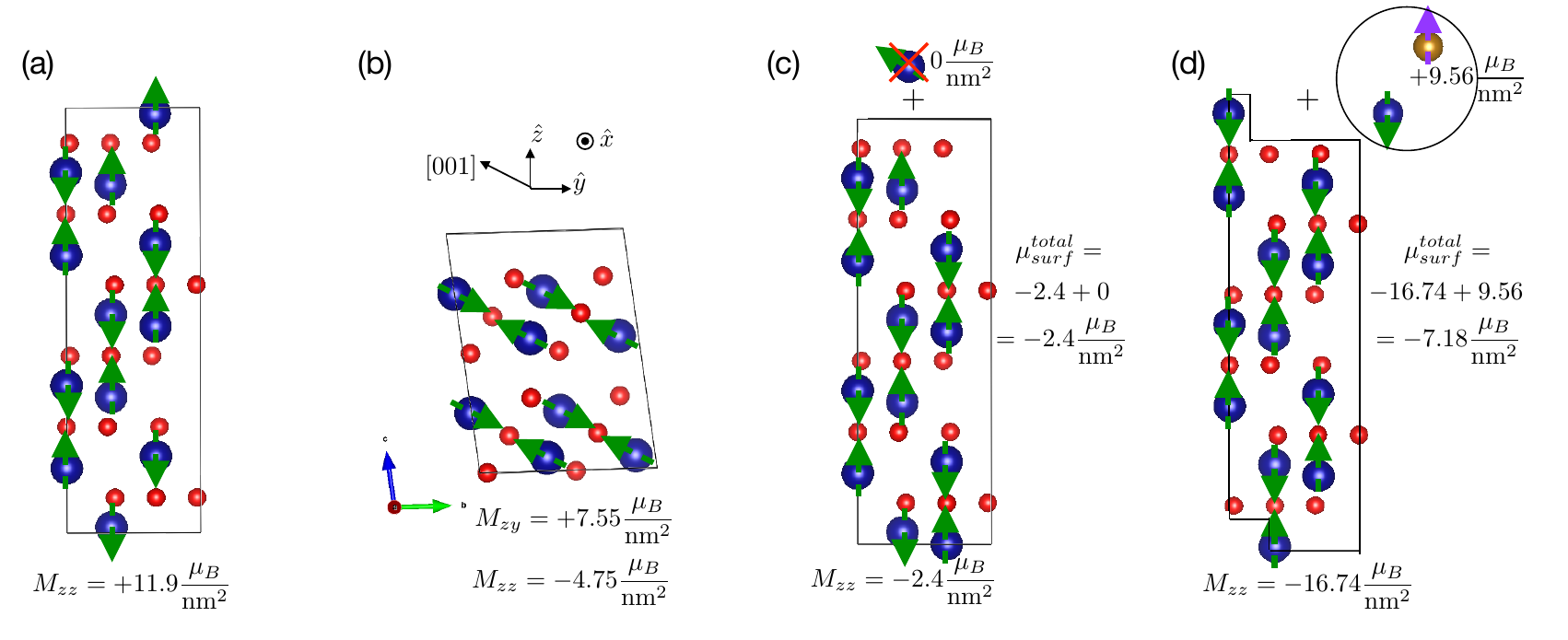}
\caption{\label{fig:mult_cell} Depiction of unit cell bases used to calculated the multipolization tensor and associated surface magnetization for the various surfaces and methods. (a), (b), (c) and (d) correspond to first, second, third and four unit cells in Table \ref{tab:cell_pos} respectively. Green arrows depict the $3$ $\mathrm{\mu_B}$ $\mathrm{Cr^{3+}}$ magnetic moments, and the purple arrow indicates the formally $5$ $\mathrm{\mu_B}$ $\mathrm{Fe^{3+}}$ magnetic moment. (a) $(001)$ pristine $\mathrm{Cr_2O_3}$ surface, assuming the top-most $\mathrm{Cr}$ is magnetically ordered. (b) $(\bar{1}02)$ surface. (c) Realistic $(001)$ $\mathrm{Cr_2O_3}$ surface, discussed in the main text, where the topmost $\mathrm{Cr}$ layer is paramagnetic. Here, the multipolizaiton tensor is calculated using a basis which can be periodically tiled, and then no additional contribution is added from the paramagnetic layer, giving a total surface magnetization $\mu_{surf}^{total}$ which is the equal to $\mathcal{M}_{zz}$ from the bulk unit cell. (d) Unit cell corresponding to bulk periodic part of $(001)$ $\mathrm{Cr_2O_3}$, plus contribution from nonperiodic $\mathrm{Fe}$ and flipped $\mathrm{Cr}$ near the surface. The magnetic moments for the nonperiodic part divided by the cross-sectional area of the $(001)$ unit cell give $+9.56 \mu_B/\mathrm{nm^2}$, which is added to the multipolization $\mathcal{M}_{zz}=-16.74\mu_B/\mathrm{nm^2}$ From the bulk unit cell to give a total estimated $\mu_{surf}^{total}$ of $+7.18\mu_B/\mathrm{nm^2}$ along $[001]$.} 
\end{figure*}
\begin{table*}
\caption{Positions, in fractional coordinates, and corresponding magnetic moments, all oriented along $[001]$, for magnetic ions in the four unit cells shown in Figure \ref{fig:mult_cell}, used to calculate the multipolization tensors. Cartesian components, in $\mathrm{\AA}$, for  the lattice vectors of the hexagonal $[001]$ oriented cells are $\mathbf{a}=[2.46,-4.26,0.0]$, $\mathbf{b}=[2.46,4.26,0.0]$ and $\mathbf{c}=[0,0,13.53]$. For the $[001]$ oriented surfaces, the moments only have a $\hat{z}$ component as $\hat{z}\parallel[001]$. The lattice vectors for the monoclinic $[\bar{1}02]$ oriented cell are $\mathbf{a}=[2.46,-4.26,0.0]$, $\mathbf{b}=[-2.46,1.42,-4.51]$ and $\mathbf{c}=[-4.92,2.84,4.51]$ with the standard $\hat{z}\parallel [001]$ Cartesian basis. They are $\mathbf{a}=[4.92,0.0,0.0]$, $\mathbf{b}=[0.0,5.33,0.0]$ and $\mathbf{c}=[0.0,-0.79,7.21]$ in the rotated Cartesian basis where $\hat{z}\parallel[\bar{1}02]$. The $\hat{x}$, $\hat{y}$, and $\hat{z}$ components of the $[001]$-oriented magnetic moments are given in this rotated basis.}\label{tab:cell_pos} 
	\begin{tabular}{| c | c | c | c | c | c | c | c | c | c | c | c | c | c | c | c | c | c | c |}
 \hline
&  \multicolumn{4}{|c|}{ordered $(001)$ surface} 
&
 \multicolumn{6}{|c|}{ ($(\bar{1}02)$ pristine surface}
 & 
 \multicolumn{4}{|c|}{paramagnetic $(001)$ surface}
 & 
 \multicolumn{4}{|c|}{$(001)$ surface with $\mathrm{Fe}$ monolayer}\\
 \hline
$\mathrm{Cr}$ site & $r^{\mathbf{a}}$ &  $r^{\mathbf{b}}$ &  $r^{\mathbf{c}}$ & $\mu^z$ & $r^{\mathbf{a}}$ &  $r^{\mathbf{b}}$ &  $r^{\mathbf{c}}$ & $\mu^x$ & $\mu^y$ & $\mu^z$ & $r^{\mathbf{a}}$ &  $r^{\mathbf{b}}$ &  $r^{\mathbf{c}}$ & $\mu^z$ & $r^{\mathbf{a}}$ &  $r^{\mathbf{b}}$ &  $r^{\mathbf{c}}$ & $\mu^z$\\
\hline
1 & 0.33 & 0.67 & 0.014 & -3 & 0.694 & 0.389 & 0.056 & 0 & -2.54 & +1.60 & 0.67 & 0.33 & 0.00 & +3 & 0.33 & 0.67 & 0.00 & +3  \\ \hline
2 &  0.00 & 0.00 & 0.153 & +3 & 0.194 & 0.889 & 0.056 & 0 & -2.54 & +1.60& 0.33 & 0.67 & 0.028 & -3 & 0.67 & 0.33 & 0.167 & +3 \\ \hline
3 & 0.67 & 0.33 & 0.181 & -3 & 0.194 & 0.50 & 0.25 & 0 & 2.54 & -1.60 & 0.00 & 0.00 & 0.167 & +3 & 0.33 & 0.67 & 0.195 & -3 \\ \hline
4 &  0.33 & 0.67 & 0.319 & +3 & 0.694 & 0.00 & 0.25 &  0 & 2.54 & -1.60&0.67 & 0.33 & 0.195 & -3 & 0.00 & 0.00 & 0.33 & +3 \\ \hline
5 &  0.00 & 0.00 & 0.347 & -3& 0.694 & 0.389 & 0.56 & 0 & -2.54 & +1.60 &0.33 & 0.67 & 0.333 & +3 & 0.67 & 0.33 & 0.361 & -3 \\ \hline
6 &  0.67 & 0.33 & 0.486 & +3 & 0.194 & 0.889 & 0.56 & 0 & -2.54 & +1.60 & 0.00 & 0.00 & 0.361 & -3 & 0.33 & 0.67 & 0.50 & +3 \\ \hline
7 &  0.33 & 0.67 & 0.514 & -3 & 0.194 & 0.50 & 0.75 & 0 & 2.54 & -1.60 & 0.67 & 0.33 & 0.50 & +3 & 0.00 & 0.00 & 0.528 & -3 \\ \hline
8 & 0.0 & 0.0 & 0.653 & +3 & 0.694 & 0.00 & 0.75 & 0 & 2.54 & -1.60 & 0.33 & 0.67 & 0.528 & -3  & 0.67 & 0.33 & 0.667 & +3 \\ \hline
9 &  0.67 & 0.33 & 0.681 & -3 & - & - & - & - & - & - & 0.00 & 0.00 & 0.667 & +3 & 0.33 & 0.67 & 0.695 & -3  \\ \hline
10 &  0.33 & 0.67 & 0.819 & +3  & - & - & - & - & - & - & 0.67 & 0.33 & 0.695 & -3 &0.00 & 0.00 & 0.833 & +3\\ \hline
11 & 0.00 & 0.00 & 0.847 & -3 & - & - & - & - & - & - & 0.33 & 0.67 & 0.833 & +3 &0.67 & 0.33 & 0.861 & -3 \\ \hline
12 & 0.67 & 0.33 & 0.986 & +3 & - & - & - & - & - & - & 0.00 & 0.00 & 0.861 & -3 & 0.00 & 0.00 & 1.028 & -3 \\ \hline
\end{tabular}
\end{table*}
In Table \ref{tab:cell_pos} We give the positions (in fractional coordinates) of the magnetic ions, as well as the magnetic moments, in the unit cell bases which are used to calculate the multipolization tensors corresponding to magnetization on the $(001)$ and ($\bar{1}02)$ surfaces of $\mathrm{Cr_2O_3}$. Recall that with a local moment approximation, the multipolization tensor $\mathcal{M}$ can be calculated as:
\begin{equation}
\mathcal{M}_{ij}^{lm}=\frac{1}{V}\sum_{\alpha}R_{i,\alpha}m_{j,\alpha},
\label{eq:M_lm}
\end{equation}
where $V$ is the unit cell volume, and the sum is over magnetic atoms in the unit cell. Equation \ref{eq:M_lm} is equivalent to Equation 1 in the main manuscript. By inspection of the form of $\mathcal{M}$ it is clear that the $ij^{\mathrm{th}}$ component of the tensor should ideally correspond to the component of magnetization which is oriented along $\hat{i}$ on a surface whose normal is parallel to $\hat{j}$. We remind the reader however that for a given Miller plane with its nonpolar termination, only the out-of-plane dimension of the corresponding bulk unit cell is unambiguously determined; each in-plane lattice constant for the unit cell used to calculate $\mathcal{M}$ can correspond to any arbitrary branch of multipolization increment which is parallel to the surface normal. Therefore, to reliably calculate the three cartesian components of magnetization on a given surface, one should calculate $\mathcal{M}$ within a rotated basis where the $\hat{z}$ cartesian axis is parallel to the surface normal. This is already the case for the $(001)$ surface with the conventional hexagonal unit cell. To obtain $\mathcal{M}$ for the $(\bar{1}02)$ surface, we rewrite the lattice vectors in a rotated cartesian coordinate system with $\hat{z}\parallel \hat{\mathbf{n}}$ where $\hat{\mathbf{n}}$ is the surface normal $[\bar{1},0,2]$, and $\hat{x}$ is parallel to the in-plane $\mathbf{a}$ lattice vector; The lattice vectors in this rotated basis are also given in the table caption. The $\hat{x}$, $\hat{y}$, and $\hat{z}$ components of the $[001]$-oriented $3\mu_B$ $\mathrm{Cr}$ magnetic moments are obtained simply by taking the dot product of the $[001]$ direction with the rotated cartesian vectors. From this we predict an in-plane (out-of-plane) $(\bar{1}02)$ surface magnetization of $+7.55$ ($-4.75$) $\mu_B/\mathrm{nm^2}$ respectively as stated in the manuscript. The in-plane magnetization is fully along the rotated cartesian $\hat{y}$ direction corresponding to a nonzero $\mathcal{M}_{zy}$ component. The corresponding bulk unit cell and basis for the $(\bar{1}02)$ surface is also depicted in Figure \ref{fig:mult_cell}(b).\\
\indent We point out here that the surface magnetization is only rigorously tied to the bulk multipolization tensor in the absence of any surface reconstruction, spin disorder, spin flipping, or doping; if the surface is truly just an abrupt termination of the bulk, the surface magnetization can then be determined by calculating $\mathcal{M}$ for the bulk unit cell which periodically tiles the semi-infinite surface of interest completely analogously to the procedure for determining bound surface charge from $\mathbf{P}_{\mathrm{bulk}}$. This is the case for the first two columns of Table \ref{tab:cell_pos}, representing respectively the $(001)$ surface assuming the surface $\mathrm{Cr}$ retain the bulk AFM order (corresponding to the basis in Figure \ref{fig:mult_cell}(a), identical to that in the left-hand side of Figure 1(b) in the main text), and the $(\bar{1}02)$ surface. Recall from the main text that since $\lambda_{surf}\sim\lambda_{bulk}$ for the $(\bar{1}02)$ Miller plane, the outermost $\mathrm{Cr}$ have bulk-like magnetization and the mutlipolization calculated from Equation \ref{eq:M_lm} in this case corresponds rigorously to the bulk $\mathcal{M}$ and should yield the true surface magnetization. However, for the other two cases we discuss, i.e. realistic $(001)$ $\mathrm{Cr_2O_3}$ with a paramagnetic $\mathrm{Cr}$ surface layer, and $(001)$ $\mathrm{Cr_2O_3}$ with a single monolayer of $\mathrm{Fe}$ there is no bulk unit cell which can be tiled semi-infinitely to define the surface.\\
\indent One way to approximately calculate the surface magnetization in this case is the method discussed in the main manuscript for $(001)$ $\mathrm{Cr_2O_3}$ with a paramagnetic surface layer. Here, a bulk multipolization tensor is calculated based on the unit cell which can be tiled parallel to the surface normal up to where the material maintains bulk character. For pristine $(001)$ $\mathrm{Cr_2O_3}$ this corresponds to unit cell shown in Figure \ref{fig:mult_cell}(c), identical to that on the right-hand side of Figure 1(b) in the main manuscript. For $\mathrm{Fe}$-capped $\mathrm{Cr_2O_3}$ it corresponds to the non-rectangular unit cell shown in Figure \ref{fig:mult_cell}(d), which excludes the $\mathrm{Fe}$ monolayer and the $\mathrm{Cr}$ which has flipped with respect to the bulk AFM magnetic order. The corresponding positions, in direct coordinates, for the magnetic $\mathrm{Cr}$ in these units cells are also given in Table \ref{tab:cell_pos}, yielding $\mathcal{M}_{zz}$ components of $-2.4$ $\mu_B/\mathrm{nm}^2$ and $-16.7$ $\mu_B/\mathrm{nm^2}$ respectively. Then, to calculate the full theoretical magnetization for the actual surface, one adds to the multipolization tensor-based value from this periodic unit cell the remaining nonperiodic contribution. For Figure \ref{fig:mult_cell}(c) the magnetization contribution from the outermost paramagnetic layer is just zero, whereas for Figure \ref{fig:mult_cell}(d) this can be approximated by summing the magnetic moment vectors for $\mathrm{Fe}$ and the flipped $\mathrm{Cr}$ ($5-3=+2$ $\mu_B$) and dividing by the cross-sectional area of the $[001]$-oriented unit cell, yielding $+9.56$ $\mu_B/\mathrm{nm^2}$. Adding these ``nonperiodic" contributions to the $M_{zz}$ components of the multipolization tensors as calculated from equation \ref{eq:M_lm} for the bulk unit cells gives $-2.4$ $\mu_B/\mathrm{nm^2}$ (as quoted in the manuscript) and $-16.74+9.56=-7.18$ $\mu_B/\mathrm{nm^2}$ for $(001)$ pristine and $\mathrm{Fe}$-capped surfaces respectively.\\
\section{Feasibility of synthesising $\mathrm{Cr_2O_3}$ capped with an $\mathrm{Fe}$ monolayer}
To assess the feasibility of terminating a $\mathrm{Cr_2O_3}$ slab with $\mathrm{Fe}$, we have calculated the relative stability a relaxed $\mathrm{Cr_{11}Fe_1O_{36}}$ slab structure with $\mathrm{Fe}$ replacing $\mathrm{Cr}$ in the two positions directly below the topmost oxygen layer ($\mathrm{Cr}$ $11$ and $\mathrm{Cr}$ $10$ positions according to the labeling in Figure 1(a) of the main text). We estimate the liklihood of substitution of $\mathrm{Fe}$ on these sites rather than the terminating $12^{\mathrm{th}}$ position by first calculating the total energies within DFT+U for the spin-polarized slabs with $\mathrm{Fe}$ in the $11^{\mathrm{th}}$ and $\mathrm{10{th}}$ positions after fully relaxing the unit cell-thick slab. We then calculate the $\mathrm{Fe}$-$\mathrm{Cr}$ couplings for the structure with $\mathrm{Fe}$ in these intermediate positions via the usual total energy method described earlier. We next calculate the Heisenberg contribution to the total DFT+U total energy using these $\mathrm{Fe}$-$\mathrm{Cr}$ $J$ values (as well as the $\mathrm{Cr}$-$\mathrm{Cr}$ $J$s for the $\mathrm{Cr}$ in the unit cell)  along with the relative directions of the magnetic moments in the DFT+U calculation. Finally, we subtract off the magnetic Heisenberg contribution from the DFT+U total energy; because the likelihood of site substitution is primarily dependent on the atomic environment, we can get a better idea of relative formation stability by neglecting magnetic contributions to the energy. We find that the resulting energies for $\mathrm{Fe}$ in the $11^{\mathrm{th}}$ and $10^{\mathrm{th}}$ positions are $0.877$ $\mathrm{eV}$ and $0.896$ $\mathrm{eV}$ higher respectively than the structure in Figure 4(c) of the main text with a terminating $\mathrm{Fe}$ layer in the $12^{\mathrm{th}}$ position. Thus at room temperature $\mathrm{k_BT}\sim 0.025$ $\mathrm{eV}$, the probability of $\mathrm{Fe}$ substituting $\mathrm{Cr}$ at these sites is suppressed, and introducing $\mathrm{Fe}$ into a vacuum chamber at the very end of growth should lead to a reasonably uniform $\mathrm{Fe}$ monolayer on the top of $\mathrm{Cr_2O_3}$. 
 We recognize that synthesis of the final hypothetical structure discussed in the main text, with a single $\mathrm{Fe}$ terminating a $[001]$ oriented slab of $\mathrm{Cr_2O_3}$, is nontrivial, since the oxygen termination of the $(001)$ $\mathrm{Cr_2O_3}$ structure which would allow subsequent deposition of the $\mathrm{Fe}$ monolayer is polar. Nevertheless, we believe it would be feasible, particularly given several experimental studies indicated that the oxygen termination of chromia can be stabilized by varying the oxygen partial pressure during synthesis\cite{Lubbe2009,Bikondoa2010,Kaspar2013}.
\end{appendix}

\FloatBarrier
\bibliography{Cr2O3_surfmag.bib}

\begin{thebibliography}{41}%
\makeatletter
\providecommand \@ifxundefined [1]{%
 \@ifx{#1\undefined}
}%
\providecommand \@ifnum [1]{%
 \ifnum #1\expandafter \@firstoftwo
 \else \expandafter \@secondoftwo
 \fi
}%
\providecommand \@ifx [1]{%
 \ifx #1\expandafter \@firstoftwo
 \else \expandafter \@secondoftwo
 \fi
}%
\providecommand \natexlab [1]{#1}%
\providecommand \enquote  [1]{``#1''}%
\providecommand \bibnamefont  [1]{#1}%
\providecommand \bibfnamefont [1]{#1}%
\providecommand \citenamefont [1]{#1}%
\providecommand \href@noop [0]{\@secondoftwo}%
\providecommand \href [0]{\begingroup \@sanitize@url \@href}%
\providecommand \@href[1]{\@@startlink{#1}\@@href}%
\providecommand \@@href[1]{\endgroup#1\@@endlink}%
\providecommand \@sanitize@url [0]{\catcode `\\12\catcode `\$12\catcode
  `\&12\catcode `\#12\catcode `\^12\catcode `\_12\catcode `\%12\relax}%
\providecommand \@@startlink[1]{}%
\providecommand \@@endlink[0]{}%
\providecommand \url  [0]{\begingroup\@sanitize@url \@url }%
\providecommand \@url [1]{\endgroup\@href {#1}{\urlprefix }}%
\providecommand \urlprefix  [0]{URL }%
\providecommand \Eprint [0]{\href }%
\providecommand \doibase [0]{http://dx.doi.org/}%
\providecommand \selectlanguage [0]{\@gobble}%
\providecommand \bibinfo  [0]{\@secondoftwo}%
\providecommand \bibfield  [0]{\@secondoftwo}%
\providecommand \translation [1]{[#1]}%
\providecommand \BibitemOpen [0]{}%
\providecommand \bibitemStop [0]{}%
\providecommand \bibitemNoStop [0]{.\EOS\space}%
\providecommand \EOS [0]{\spacefactor3000\relax}%
\providecommand \BibitemShut  [1]{\csname bibitem#1\endcsname}%
\let\auto@bib@innerbib\@empty
\bibitem [{\citenamefont {Astrov}(1960)}]{Astrov1960}%
  \BibitemOpen
  \bibfield  {author} {\bibinfo {author} {\bibfnamefont {D.~N.}\ \bibnamefont
  {Astrov}},\ }\href@noop {} {\bibfield  {journal} {\bibinfo  {journal} {J.
  Exptl. Theoret. Phys. (U.S.S.R.)}\ }\textbf {\bibinfo {volume} {11}},\
  \bibinfo {pages} {984} (\bibinfo {year} {1960})}\BibitemShut {NoStop}%
\bibitem [{\citenamefont {Belashchenko}(2010)}]{Belashchenko2010}%
  \BibitemOpen
  \bibfield  {author} {\bibinfo {author} {\bibfnamefont {K.~D.}\ \bibnamefont
  {Belashchenko}},\ }\href {\doibase 10.1103/PhysRevLett.105.147204} {\bibfield
   {journal} {\bibinfo  {journal} {Physical Review Letters}\ }\textbf {\bibinfo
  {volume} {105}} (\bibinfo {year} {2010}),\
  10.1103/PhysRevLett.105.147204}\BibitemShut {NoStop}%
\bibitem [{\citenamefont {Ye}(2022)}]{Ye2022}%
  \BibitemOpen
  \bibfield  {author} {\bibinfo {author} {\bibfnamefont {S.}~\bibnamefont
  {Ye}},\ }\href {\doibase 10.1002/pssr.202100396} {\bibfield  {journal}
  {\bibinfo  {journal} {Physica Status Solidi - Rapid Research Letters}\
  }\textbf {\bibinfo {volume} {16}} (\bibinfo {year} {2022}),\
  10.1002/pssr.202100396}\BibitemShut {NoStop}%
\bibitem [{\citenamefont {Hedrich}\ \emph {et~al.}(2021)\citenamefont
  {Hedrich}, \citenamefont {Wagner}, \citenamefont {Pylypovskyi}, \citenamefont
  {Shields}, \citenamefont {Kosub}, \citenamefont {Sheka}, \citenamefont
  {Makarov},\ and\ \citenamefont {Maletinsky}}]{Hedrich2021}%
  \BibitemOpen
  \bibfield  {author} {\bibinfo {author} {\bibfnamefont {N.}~\bibnamefont
  {Hedrich}}, \bibinfo {author} {\bibfnamefont {K.}~\bibnamefont {Wagner}},
  \bibinfo {author} {\bibfnamefont {O.~V.}\ \bibnamefont {Pylypovskyi}},
  \bibinfo {author} {\bibfnamefont {B.~J.}\ \bibnamefont {Shields}}, \bibinfo
  {author} {\bibfnamefont {T.}~\bibnamefont {Kosub}}, \bibinfo {author}
  {\bibfnamefont {D.~D.}\ \bibnamefont {Sheka}}, \bibinfo {author}
  {\bibfnamefont {D.}~\bibnamefont {Makarov}}, \ and\ \bibinfo {author}
  {\bibfnamefont {P.}~\bibnamefont {Maletinsky}},\ }\href {\doibase
  10.1038/s41567-020-01157-0} {\bibfield  {journal} {\bibinfo  {journal}
  {Nature Physics}\ }\textbf {\bibinfo {volume} {17}},\ \bibinfo {pages} {574}
  (\bibinfo {year} {2021})}\BibitemShut {NoStop}%
\bibitem [{\citenamefont {Nogue´s}\ and\ \citenamefont
  {Schuller}(1999)}]{Nogues1999}%
  \BibitemOpen
  \bibfield  {author} {\bibinfo {author} {\bibfnamefont {J.~N.}\ \bibnamefont
  {Nogue´s}}\ and\ \bibinfo {author} {\bibfnamefont {I.~K.}\ \bibnamefont
  {Schuller}},\ }\href@noop {} {\bibfield  {journal} {\bibinfo  {journal}
  {Journal of Magnetism and Magnetic Materials}\ }\textbf {\bibinfo {volume}
  {192}},\ \bibinfo {pages} {203} (\bibinfo {year} {1999})}\BibitemShut
  {NoStop}%
\bibitem [{\citenamefont {Stamps}(2000)}]{Stamps2000}%
  \BibitemOpen
  \bibfield  {author} {\bibinfo {author} {\bibfnamefont {R.~L.}\ \bibnamefont
  {Stamps}},\ }\href@noop {} {\bibfield  {journal} {\bibinfo  {journal}
  {Journal of Physics D: Applied}\ }\textbf {\bibinfo {volume} {33}} (\bibinfo
  {year} {2000})}\BibitemShut {NoStop}%
\bibitem [{\citenamefont {Echtenkamp}(2021)}]{Echtenkamp2021}%
  \BibitemOpen
  \bibfield  {author} {\bibinfo {author} {\bibfnamefont {W.}~\bibnamefont
  {Echtenkamp}},\ }\href {https://digitalcommons.unl.edu/physicsdiss} {\enquote
  {\bibinfo {title} {Voltage-controlled magnetization in chromia-based magnetic
  heterostructures heterostructures},}\ } (\bibinfo {year} {2021})\BibitemShut
  {NoStop}%
\bibitem [{\citenamefont {Schlitz}\ \emph {et~al.}(2018)\citenamefont
  {Schlitz}, \citenamefont {Kosub}, \citenamefont {Thomas}, \citenamefont
  {Fabretti}, \citenamefont {Nielsch}, \citenamefont {Makarov},\ and\
  \citenamefont {Goennenwein}}]{Schlitz2018}%
  \BibitemOpen
  \bibfield  {author} {\bibinfo {author} {\bibfnamefont {R.}~\bibnamefont
  {Schlitz}}, \bibinfo {author} {\bibfnamefont {T.}~\bibnamefont {Kosub}},
  \bibinfo {author} {\bibfnamefont {A.}~\bibnamefont {Thomas}}, \bibinfo
  {author} {\bibfnamefont {S.}~\bibnamefont {Fabretti}}, \bibinfo {author}
  {\bibfnamefont {K.}~\bibnamefont {Nielsch}}, \bibinfo {author} {\bibfnamefont
  {D.}~\bibnamefont {Makarov}}, \ and\ \bibinfo {author} {\bibfnamefont
  {S.~T.}\ \bibnamefont {Goennenwein}},\ }\href {\doibase 10.1063/1.5019934}
  {\bibfield  {journal} {\bibinfo  {journal} {Applied Physics Letters}\
  }\textbf {\bibinfo {volume} {112}} (\bibinfo {year} {2018}),\
  10.1063/1.5019934}\BibitemShut {NoStop}%
\bibitem [{\citenamefont {Muduli}\ \emph {et~al.}(2021)\citenamefont {Muduli},
  \citenamefont {Schlitz}, \citenamefont {Kosub}, \citenamefont {Hübner},
  \citenamefont {Erbe}, \citenamefont {Makarov},\ and\ \citenamefont
  {Goennenwein}}]{Muduli2021}%
  \BibitemOpen
  \bibfield  {author} {\bibinfo {author} {\bibfnamefont {P.}~\bibnamefont
  {Muduli}}, \bibinfo {author} {\bibfnamefont {R.}~\bibnamefont {Schlitz}},
  \bibinfo {author} {\bibfnamefont {T.}~\bibnamefont {Kosub}}, \bibinfo
  {author} {\bibfnamefont {R.}~\bibnamefont {Hübner}}, \bibinfo {author}
  {\bibfnamefont {A.}~\bibnamefont {Erbe}}, \bibinfo {author} {\bibfnamefont
  {D.}~\bibnamefont {Makarov}}, \ and\ \bibinfo {author} {\bibfnamefont
  {S.~T.}\ \bibnamefont {Goennenwein}},\ }\href {\doibase 10.1063/5.0037860}
  {\bibfield  {journal} {\bibinfo  {journal} {APL Materials}\ }\textbf
  {\bibinfo {volume} {9}} (\bibinfo {year} {2021}),\
  10.1063/5.0037860}\BibitemShut {NoStop}%
\bibitem [{\citenamefont {Appel}\ \emph {et~al.}(2019)\citenamefont {Appel},
  \citenamefont {Shields}, \citenamefont {Kosub}, \citenamefont {Hedrich},
  \citenamefont {Hübner}, \citenamefont {Faßbender}, \citenamefont
  {Makarov},\ and\ \citenamefont {Maletinsky}}]{Appel2019}%
  \BibitemOpen
  \bibfield  {author} {\bibinfo {author} {\bibfnamefont {P.}~\bibnamefont
  {Appel}}, \bibinfo {author} {\bibfnamefont {B.~J.}\ \bibnamefont {Shields}},
  \bibinfo {author} {\bibfnamefont {T.}~\bibnamefont {Kosub}}, \bibinfo
  {author} {\bibfnamefont {N.}~\bibnamefont {Hedrich}}, \bibinfo {author}
  {\bibfnamefont {R.}~\bibnamefont {Hübner}}, \bibinfo {author} {\bibfnamefont
  {J.}~\bibnamefont {Faßbender}}, \bibinfo {author} {\bibfnamefont
  {D.}~\bibnamefont {Makarov}}, \ and\ \bibinfo {author} {\bibfnamefont
  {P.}~\bibnamefont {Maletinsky}},\ }\href {\doibase
  10.1021/acs.nanolett.8b04681} {\bibfield  {journal} {\bibinfo  {journal}
  {Nano Letters}\ }\textbf {\bibinfo {volume} {19}},\ \bibinfo {pages} {1682}
  (\bibinfo {year} {2019})}\BibitemShut {NoStop}%
\bibitem [{\citenamefont {Wörnle}\ \emph {et~al.}(2021)\citenamefont
  {Wörnle}, \citenamefont {Welter}, \citenamefont {Giraldo}, \citenamefont
  {Lottermoser}, \citenamefont {Fiebig}, \citenamefont {Gambardella},\ and\
  \citenamefont {Degen}}]{Wornle2021}%
  \BibitemOpen
  \bibfield  {author} {\bibinfo {author} {\bibfnamefont {M.~S.}\ \bibnamefont
  {Wörnle}}, \bibinfo {author} {\bibfnamefont {P.}~\bibnamefont {Welter}},
  \bibinfo {author} {\bibfnamefont {M.}~\bibnamefont {Giraldo}}, \bibinfo
  {author} {\bibfnamefont {T.}~\bibnamefont {Lottermoser}}, \bibinfo {author}
  {\bibfnamefont {M.}~\bibnamefont {Fiebig}}, \bibinfo {author} {\bibfnamefont
  {P.}~\bibnamefont {Gambardella}}, \ and\ \bibinfo {author} {\bibfnamefont
  {C.~L.}\ \bibnamefont {Degen}},\ }\href {\doibase
  10.1103/PhysRevB.103.094426} {\bibfield  {journal} {\bibinfo  {journal}
  {Physical Review B}\ }\textbf {\bibinfo {volume} {103}} (\bibinfo {year}
  {2021}),\ 10.1103/PhysRevB.103.094426}\BibitemShut {NoStop}%
\bibitem [{\citenamefont {Spaldin}(2021)}]{Spaldin2021}%
  \BibitemOpen
  \bibfield  {author} {\bibinfo {author} {\bibfnamefont {N.~A.}\ \bibnamefont
  {Spaldin}},\ }\href {\doibase 10.1134/S1063776121040208} {\bibfield
  {journal} {\bibinfo  {journal} {Journal of Experimental and Theoretical
  Physics}\ }\textbf {\bibinfo {volume} {132}},\ \bibinfo {pages} {493}
  (\bibinfo {year} {2021})}\BibitemShut {NoStop}%
\bibitem [{\citenamefont {Stengel}(2011)}]{Stengel2011}%
  \BibitemOpen
  \bibfield  {author} {\bibinfo {author} {\bibfnamefont {M.}~\bibnamefont
  {Stengel}},\ }\href {\doibase 10.1103/PhysRevB.84.205432} {\bibfield
  {journal} {\bibinfo  {journal} {Physical Review B - Condensed Matter and
  Materials Physics}\ }\textbf {\bibinfo {volume} {84}} (\bibinfo {year}
  {2011}),\ 10.1103/PhysRevB.84.205432}\BibitemShut {NoStop}%
\bibitem [{\citenamefont {Nakagawa}\ \emph {et~al.}(2006)\citenamefont
  {Nakagawa}, \citenamefont {Hwang},\ and\ \citenamefont
  {Muller}}]{Nakagawa2006}%
  \BibitemOpen
  \bibfield  {author} {\bibinfo {author} {\bibfnamefont {N.}~\bibnamefont
  {Nakagawa}}, \bibinfo {author} {\bibfnamefont {H.~Y.}\ \bibnamefont {Hwang}},
  \ and\ \bibinfo {author} {\bibfnamefont {D.~A.}\ \bibnamefont {Muller}},\
  }\href {\doibase 10.1038/nmat1569} {\bibfield  {journal} {\bibinfo  {journal}
  {Nature Materials}\ }\textbf {\bibinfo {volume} {5}},\ \bibinfo {pages} {204}
  (\bibinfo {year} {2006})}\BibitemShut {NoStop}%
\bibitem [{\citenamefont {Vanderbilt}\ and\ \citenamefont
  {King-Smith}(1993)}]{King-Smith1993b}%
  \BibitemOpen
  \bibfield  {author} {\bibinfo {author} {\bibfnamefont {D.}~\bibnamefont
  {Vanderbilt}}\ and\ \bibinfo {author} {\bibfnamefont {R.~D.}\ \bibnamefont
  {King-Smith}},\ }\href@noop {} {\bibfield  {journal} {\bibinfo  {journal}
  {Physical Review B}\ }\textbf {\bibinfo {volume} {48}},\ \bibinfo {pages}
  {4442} (\bibinfo {year} {1993})}\BibitemShut {NoStop}%
\bibitem [{\citenamefont {King-Smith}\ and\ \citenamefont
  {Vanderbilt}(1993)}]{King-Smith1993}%
  \BibitemOpen
  \bibfield  {author} {\bibinfo {author} {\bibfnamefont {R.~D.}\ \bibnamefont
  {King-Smith}}\ and\ \bibinfo {author} {\bibfnamefont {D.}~\bibnamefont
  {Vanderbilt}},\ }\href@noop {} {\bibfield  {journal} {\bibinfo  {journal}
  {Physical Review B}\ }\textbf {\bibinfo {volume} {47}},\ \bibinfo {pages}
  {1651} (\bibinfo {year} {1993})}\BibitemShut {NoStop}%
\bibitem [{\citenamefont {Spaldin}\ \emph {et~al.}(2013)\citenamefont
  {Spaldin}, \citenamefont {Fechner}, \citenamefont {Bousquet}, \citenamefont
  {Balatsky},\ and\ \citenamefont {Nordström}}]{Spaldin2013}%
  \BibitemOpen
  \bibfield  {author} {\bibinfo {author} {\bibfnamefont {N.~A.}\ \bibnamefont
  {Spaldin}}, \bibinfo {author} {\bibfnamefont {M.}~\bibnamefont {Fechner}},
  \bibinfo {author} {\bibfnamefont {E.}~\bibnamefont {Bousquet}}, \bibinfo
  {author} {\bibfnamefont {A.}~\bibnamefont {Balatsky}}, \ and\ \bibinfo
  {author} {\bibfnamefont {L.}~\bibnamefont {Nordström}},\ }\href {\doibase
  10.1103/PhysRevB.88.094429} {\bibfield  {journal} {\bibinfo  {journal}
  {Physical Review B - Condensed Matter and Materials Physics}\ }\textbf
  {\bibinfo {volume} {88}} (\bibinfo {year} {2013}),\
  10.1103/PhysRevB.88.094429}\BibitemShut {NoStop}%
\bibitem [{\citenamefont {He}\ \emph {et~al.}(2010)\citenamefont {He},
  \citenamefont {Wang}, \citenamefont {Wu}, \citenamefont {Caruso},
  \citenamefont {Vescovo}, \citenamefont {Belashchenko}, \citenamefont
  {Dowben},\ and\ \citenamefont {Binek}}]{He2010}%
  \BibitemOpen
  \bibfield  {author} {\bibinfo {author} {\bibfnamefont {X.}~\bibnamefont
  {He}}, \bibinfo {author} {\bibfnamefont {Y.}~\bibnamefont {Wang}}, \bibinfo
  {author} {\bibfnamefont {N.}~\bibnamefont {Wu}}, \bibinfo {author}
  {\bibfnamefont {A.~N.}\ \bibnamefont {Caruso}}, \bibinfo {author}
  {\bibfnamefont {E.}~\bibnamefont {Vescovo}}, \bibinfo {author} {\bibfnamefont
  {K.~D.}\ \bibnamefont {Belashchenko}}, \bibinfo {author} {\bibfnamefont
  {P.~A.}\ \bibnamefont {Dowben}}, \ and\ \bibinfo {author} {\bibfnamefont
  {C.}~\bibnamefont {Binek}},\ }\href {\doibase 10.1038/nmat2785} {\bibfield
  {journal} {\bibinfo  {journal} {Nature Materials}\ }\textbf {\bibinfo
  {volume} {9}},\ \bibinfo {pages} {579} (\bibinfo {year} {2010})}\BibitemShut
  {NoStop}%
\bibitem [{\citenamefont {Fechner}\ \emph {et~al.}(2018)\citenamefont
  {Fechner}, \citenamefont {Sukhov}, \citenamefont {Chotorlishvili},
  \citenamefont {Kenel}, \citenamefont {Berakdar},\ and\ \citenamefont
  {Spaldin}}]{Fechner2018}%
  \BibitemOpen
  \bibfield  {author} {\bibinfo {author} {\bibfnamefont {M.}~\bibnamefont
  {Fechner}}, \bibinfo {author} {\bibfnamefont {A.}~\bibnamefont {Sukhov}},
  \bibinfo {author} {\bibfnamefont {L.}~\bibnamefont {Chotorlishvili}},
  \bibinfo {author} {\bibfnamefont {C.}~\bibnamefont {Kenel}}, \bibinfo
  {author} {\bibfnamefont {J.}~\bibnamefont {Berakdar}}, \ and\ \bibinfo
  {author} {\bibfnamefont {N.~A.}\ \bibnamefont {Spaldin}},\ }\href {\doibase
  10.1103/PhysRevMaterials.2.064401} {\bibfield  {journal} {\bibinfo  {journal}
  {Physical Review Materials}\ }\textbf {\bibinfo {volume} {2}} (\bibinfo
  {year} {2018}),\ 10.1103/PhysRevMaterials.2.064401}\BibitemShut {NoStop}%
\bibitem [{\citenamefont {Samuelsent}\ \emph {et~al.}(1970)\citenamefont
  {Samuelsent}, \citenamefont {Hutchings},\ and\ \citenamefont
  {Shirane}}]{Samuelsent1970}%
  \BibitemOpen
  \bibfield  {author} {\bibinfo {author} {\bibfnamefont {E.~J.}\ \bibnamefont
  {Samuelsent}}, \bibinfo {author} {\bibfnamefont {M.~T.}\ \bibnamefont
  {Hutchings}}, \ and\ \bibinfo {author} {\bibfnamefont {G.}~\bibnamefont
  {Shirane}},\ }\href@noop {} {\bibfield  {journal} {\bibinfo  {journal}
  {Physica}\ }\textbf {\bibinfo {volume} {48}} (\bibinfo {year}
  {1970})}\BibitemShut {NoStop}%
\bibitem [{\citenamefont {Shi}\ \emph {et~al.}(2009)\citenamefont {Shi},
  \citenamefont {Wysocki},\ and\ \citenamefont {Belashchenko}}]{Shi2009}%
  \BibitemOpen
  \bibfield  {author} {\bibinfo {author} {\bibfnamefont {S.}~\bibnamefont
  {Shi}}, \bibinfo {author} {\bibfnamefont {A.~L.}\ \bibnamefont {Wysocki}}, \
  and\ \bibinfo {author} {\bibfnamefont {K.~D.}\ \bibnamefont {Belashchenko}},\
  }\href {\doibase 10.1103/PhysRevB.79.104404} {\bibfield  {journal} {\bibinfo
  {journal} {Physical Review B - Condensed Matter and Materials Physics}\
  }\textbf {\bibinfo {volume} {79}} (\bibinfo {year} {2009}),\
  10.1103/PhysRevB.79.104404}\BibitemShut {NoStop}%
\bibitem [{\citenamefont {Xiang}\ \emph {et~al.}(2011)\citenamefont {Xiang},
  \citenamefont {Kan}, \citenamefont {Wei}, \citenamefont {Whangbo},\ and\
  \citenamefont {Gong}}]{Xiang2011}%
  \BibitemOpen
  \bibfield  {author} {\bibinfo {author} {\bibfnamefont {H.~J.}\ \bibnamefont
  {Xiang}}, \bibinfo {author} {\bibfnamefont {E.~J.}\ \bibnamefont {Kan}},
  \bibinfo {author} {\bibfnamefont {S.~H.}\ \bibnamefont {Wei}}, \bibinfo
  {author} {\bibfnamefont {M.~H.}\ \bibnamefont {Whangbo}}, \ and\ \bibinfo
  {author} {\bibfnamefont {X.~G.}\ \bibnamefont {Gong}},\ }\href {\doibase
  10.1103/PhysRevB.84.224429} {\bibfield  {journal} {\bibinfo  {journal}
  {Physical Review B - Condensed Matter and Materials Physics}\ }\textbf
  {\bibinfo {volume} {84}} (\bibinfo {year} {2011}),\
  10.1103/PhysRevB.84.224429}\BibitemShut {NoStop}%
\bibitem [{\citenamefont {Kresse}\ and\ \citenamefont
  {Furthmüller}(1996)}]{Kresse1996}%
  \BibitemOpen
  \bibfield  {author} {\bibinfo {author} {\bibfnamefont {G.}~\bibnamefont
  {Kresse}}\ and\ \bibinfo {author} {\bibfnamefont {J.}~\bibnamefont
  {Furthmüller}},\ }\href@noop {} {\bibfield  {journal} {\bibinfo  {journal}
  {Physical Review B}\ }\textbf {\bibinfo {volume} {54}} (\bibinfo {year}
  {1996})}\BibitemShut {NoStop}%
\bibitem [{\citenamefont {Urru}\ and\ \citenamefont
  {Spaldin}(2022)}]{Urru2022}%
  \BibitemOpen
  \bibfield  {author} {\bibinfo {author} {\bibfnamefont {A.}~\bibnamefont
  {Urru}}\ and\ \bibinfo {author} {\bibfnamefont {N.~A.}\ \bibnamefont
  {Spaldin}},\ }\href {http://arxiv.org/abs/2206.00522} {\bibfield  {journal}
  {\bibinfo  {journal} {arXiv:2206.00522}\ } (\bibinfo {year}
  {2022})}\BibitemShut {NoStop}%
\bibitem [{\citenamefont {Mostovoy}\ \emph {et~al.}(2010)\citenamefont
  {Mostovoy}, \citenamefont {Scaramucci}, \citenamefont {Spaldin},\ and\
  \citenamefont {Delaney}}]{Mostovoy2010}%
  \BibitemOpen
  \bibfield  {author} {\bibinfo {author} {\bibfnamefont {M.}~\bibnamefont
  {Mostovoy}}, \bibinfo {author} {\bibfnamefont {A.}~\bibnamefont
  {Scaramucci}}, \bibinfo {author} {\bibfnamefont {N.~A.}\ \bibnamefont
  {Spaldin}}, \ and\ \bibinfo {author} {\bibfnamefont {K.~T.}\ \bibnamefont
  {Delaney}},\ }\href {\doibase 10.1103/PhysRevLett.105.087202} {\bibfield
  {journal} {\bibinfo  {journal} {Physical Review Letters}\ }\textbf {\bibinfo
  {volume} {105}} (\bibinfo {year} {2010}),\
  10.1103/PhysRevLett.105.087202}\BibitemShut {NoStop}%
\bibitem [{\citenamefont {Skubic}\ \emph {et~al.}(2008)\citenamefont {Skubic},
  \citenamefont {Hellsvik}, \citenamefont {Nordström},\ and\ \citenamefont
  {Eriksson}}]{Skubic2008}%
  \BibitemOpen
  \bibfield  {author} {\bibinfo {author} {\bibfnamefont {B.}~\bibnamefont
  {Skubic}}, \bibinfo {author} {\bibfnamefont {J.}~\bibnamefont {Hellsvik}},
  \bibinfo {author} {\bibfnamefont {L.}~\bibnamefont {Nordström}}, \ and\
  \bibinfo {author} {\bibfnamefont {O.}~\bibnamefont {Eriksson}},\ }\href
  {\doibase 10.1088/0953-8984/20/31/315203} {\bibfield  {journal} {\bibinfo
  {journal} {Journal of Physics Condensed Matter}\ }\textbf {\bibinfo {volume}
  {20}} (\bibinfo {year} {2008}),\ 10.1088/0953-8984/20/31/315203}\BibitemShut
  {NoStop}%
\bibitem [{\citenamefont {Wysocki}\ \emph {et~al.}(2012)\citenamefont
  {Wysocki}, \citenamefont {Shi},\ and\ \citenamefont
  {Belashchenko}}]{Wysocki2012}%
  \BibitemOpen
  \bibfield  {author} {\bibinfo {author} {\bibfnamefont {A.~L.}\ \bibnamefont
  {Wysocki}}, \bibinfo {author} {\bibfnamefont {S.}~\bibnamefont {Shi}}, \ and\
  \bibinfo {author} {\bibfnamefont {K.~D.}\ \bibnamefont {Belashchenko}},\
  }\href {\doibase 10.1103/PhysRevB.86.165443} {\bibfield  {journal} {\bibinfo
  {journal} {Physical Review B - Condensed Matter and Materials Physics}\
  }\textbf {\bibinfo {volume} {86}} (\bibinfo {year} {2012}),\
  10.1103/PhysRevB.86.165443}\BibitemShut {NoStop}%
\bibitem [{\citenamefont {Kota}\ \emph {et~al.}(2013)\citenamefont {Kota},
  \citenamefont {Imamura},\ and\ \citenamefont {Sasaki}}]{Kota2013}%
  \BibitemOpen
  \bibfield  {author} {\bibinfo {author} {\bibfnamefont {Y.}~\bibnamefont
  {Kota}}, \bibinfo {author} {\bibfnamefont {H.}~\bibnamefont {Imamura}}, \
  and\ \bibinfo {author} {\bibfnamefont {M.}~\bibnamefont {Sasaki}},\ }\href
  {\doibase 10.7567/APEX.6.113007} {\bibfield  {journal} {\bibinfo  {journal}
  {Applied Physics Express}\ }\textbf {\bibinfo {volume} {6}} (\bibinfo {year}
  {2013}),\ 10.7567/APEX.6.113007}\BibitemShut {NoStop}%
\bibitem [{\citenamefont {Borisov}\ \emph {et~al.}(2005)\citenamefont
  {Borisov}, \citenamefont {Hochstrat}, \citenamefont {Chen}, \citenamefont
  {Kleemann},\ and\ \citenamefont {Binek}}]{Borisov2005}%
  \BibitemOpen
  \bibfield  {author} {\bibinfo {author} {\bibfnamefont {P.}~\bibnamefont
  {Borisov}}, \bibinfo {author} {\bibfnamefont {A.}~\bibnamefont {Hochstrat}},
  \bibinfo {author} {\bibfnamefont {X.}~\bibnamefont {Chen}}, \bibinfo {author}
  {\bibfnamefont {W.}~\bibnamefont {Kleemann}}, \ and\ \bibinfo {author}
  {\bibfnamefont {C.}~\bibnamefont {Binek}},\ }\href {\doibase
  10.1103/PhysRevLett.94.117203} {\bibfield  {journal} {\bibinfo  {journal}
  {Physical Review Letters}\ }\textbf {\bibinfo {volume} {94}} (\bibinfo {year}
  {2005}),\ 10.1103/PhysRevLett.94.117203}\BibitemShut {NoStop}%
\bibitem [{\citenamefont {Kota}\ \emph {et~al.}(2014)\citenamefont {Kota},
  \citenamefont {Imamura},\ and\ \citenamefont {Sasaki}}]{Kota2014}%
  \BibitemOpen
  \bibfield  {author} {\bibinfo {author} {\bibfnamefont {Y.}~\bibnamefont
  {Kota}}, \bibinfo {author} {\bibfnamefont {H.}~\bibnamefont {Imamura}}, \
  and\ \bibinfo {author} {\bibfnamefont {M.}~\bibnamefont {Sasaki}},\ }\href
  {\doibase 10.1109/TMAG.2014.2324014} {\bibfield  {journal} {\bibinfo
  {journal} {IEEE Transactions on Magnetics}\ }\textbf {\bibinfo {volume} {50}}
  (\bibinfo {year} {2014}),\ 10.1109/TMAG.2014.2324014}\BibitemShut {NoStop}%
\bibitem [{\citenamefont {Dedkov}\ \emph {et~al.}(2007)\citenamefont {Dedkov},
  \citenamefont {Laubschat}, \citenamefont {Khmelevskyi}, \citenamefont
  {Redinger}, \citenamefont {Mohn},\ and\ \citenamefont
  {Weinert}}]{Dedkov2007}%
  \BibitemOpen
  \bibfield  {author} {\bibinfo {author} {\bibfnamefont {Y.~S.}\ \bibnamefont
  {Dedkov}}, \bibinfo {author} {\bibfnamefont {C.}~\bibnamefont {Laubschat}},
  \bibinfo {author} {\bibfnamefont {S.}~\bibnamefont {Khmelevskyi}}, \bibinfo
  {author} {\bibfnamefont {J.}~\bibnamefont {Redinger}}, \bibinfo {author}
  {\bibfnamefont {P.}~\bibnamefont {Mohn}}, \ and\ \bibinfo {author}
  {\bibfnamefont {M.}~\bibnamefont {Weinert}},\ }\href {\doibase
  10.1103/PhysRevLett.99.047204} {\bibfield  {journal} {\bibinfo  {journal}
  {Physical Review Letters}\ }\textbf {\bibinfo {volume} {99}} (\bibinfo {year}
  {2007}),\ 10.1103/PhysRevLett.99.047204}\BibitemShut {NoStop}%
\bibitem [{\citenamefont {Rosenberg}\ and\ \citenamefont
  {Franz}(2012)}]{Rosenberg2012}%
  \BibitemOpen
  \bibfield  {author} {\bibinfo {author} {\bibfnamefont {G.}~\bibnamefont
  {Rosenberg}}\ and\ \bibinfo {author} {\bibfnamefont {M.}~\bibnamefont
  {Franz}},\ }\href {\doibase 10.1103/PhysRevB.85.195119} {\bibfield  {journal}
  {\bibinfo  {journal} {Physical Review B - Condensed Matter and Materials
  Physics}\ }\textbf {\bibinfo {volume} {85}} (\bibinfo {year} {2012}),\
  10.1103/PhysRevB.85.195119}\BibitemShut {NoStop}%
\bibitem [{\citenamefont {Schulz}\ \emph {et~al.}(2019)\citenamefont {Schulz},
  \citenamefont {Nechaev}, \citenamefont {Güttler}, \citenamefont {Poelchen},
  \citenamefont {Generalov}, \citenamefont {Danzenbächer}, \citenamefont
  {Chikina}, \citenamefont {Seiro}, \citenamefont {Kliemt}, \citenamefont
  {Vyazovskaya}, \citenamefont {Kim}, \citenamefont {Dudin}, \citenamefont
  {Chulkov}, \citenamefont {Laubschat}, \citenamefont {Krasovskii},
  \citenamefont {Geibel}, \citenamefont {Krellner}, \citenamefont {Kummer},\
  and\ \citenamefont {Vyalikh}}]{Schulz2019}%
  \BibitemOpen
  \bibfield  {author} {\bibinfo {author} {\bibfnamefont {S.}~\bibnamefont
  {Schulz}}, \bibinfo {author} {\bibfnamefont {I.~A.}\ \bibnamefont {Nechaev}},
  \bibinfo {author} {\bibfnamefont {M.}~\bibnamefont {Güttler}}, \bibinfo
  {author} {\bibfnamefont {G.}~\bibnamefont {Poelchen}}, \bibinfo {author}
  {\bibfnamefont {A.}~\bibnamefont {Generalov}}, \bibinfo {author}
  {\bibfnamefont {S.}~\bibnamefont {Danzenbächer}}, \bibinfo {author}
  {\bibfnamefont {A.}~\bibnamefont {Chikina}}, \bibinfo {author} {\bibfnamefont
  {S.}~\bibnamefont {Seiro}}, \bibinfo {author} {\bibfnamefont
  {K.}~\bibnamefont {Kliemt}}, \bibinfo {author} {\bibfnamefont {A.~Y.}\
  \bibnamefont {Vyazovskaya}}, \bibinfo {author} {\bibfnamefont {T.~K.}\
  \bibnamefont {Kim}}, \bibinfo {author} {\bibfnamefont {P.}~\bibnamefont
  {Dudin}}, \bibinfo {author} {\bibfnamefont {E.~V.}\ \bibnamefont {Chulkov}},
  \bibinfo {author} {\bibfnamefont {C.}~\bibnamefont {Laubschat}}, \bibinfo
  {author} {\bibfnamefont {E.~E.}\ \bibnamefont {Krasovskii}}, \bibinfo
  {author} {\bibfnamefont {C.}~\bibnamefont {Geibel}}, \bibinfo {author}
  {\bibfnamefont {C.}~\bibnamefont {Krellner}}, \bibinfo {author}
  {\bibfnamefont {K.}~\bibnamefont {Kummer}}, \ and\ \bibinfo {author}
  {\bibfnamefont {D.~V.}\ \bibnamefont {Vyalikh}},\ }\href {\doibase
  10.1038/s41535-019-0166-z} {\bibfield  {journal} {\bibinfo  {journal} {npj
  Quantum Materials}\ }\textbf {\bibinfo {volume} {4}} (\bibinfo {year}
  {2019}),\ 10.1038/s41535-019-0166-z}\BibitemShut {NoStop}%
\bibitem [{\citenamefont {Blochl}(1994)}]{Blochl1994a}%
  \BibitemOpen
  \bibfield  {author} {\bibinfo {author} {\bibfnamefont {P.~E.}\ \bibnamefont
  {Blochl}},\ }\href@noop {} {\bibfield  {journal} {\bibinfo  {journal}
  {Physical Review B}\ }\textbf {\bibinfo {volume} {50}},\ \bibinfo {pages}
  {24} (\bibinfo {year} {1994})}\BibitemShut {NoStop}%
\bibitem [{\citenamefont {Anisimov}\ \emph {et~al.}(1997)\citenamefont
  {Anisimov}, \citenamefont {Aryasetiawan},\ and\ \citenamefont
  {Lichtenstein}}]{Anisimov1997}%
  \BibitemOpen
  \bibfield  {author} {\bibinfo {author} {\bibfnamefont {V.~I.}\ \bibnamefont
  {Anisimov}}, \bibinfo {author} {\bibfnamefont {F.}~\bibnamefont
  {Aryasetiawan}}, \ and\ \bibinfo {author} {\bibfnamefont {A.~I.}\
  \bibnamefont {Lichtenstein}},\ }\href@noop {} {\bibfield  {journal} {\bibinfo
   {journal} {J. Phys.: Condens. Matter}\ }\textbf {\bibinfo {volume} {9}},\
  \bibinfo {pages} {76603} (\bibinfo {year} {1997})}\BibitemShut {NoStop}%
\bibitem [{\citenamefont {Dudarev}\ \emph {et~al.}(1998)\citenamefont
  {Dudarev}, \citenamefont {Botton}, \citenamefont {Savrasov}, \citenamefont
  {Humphreys},\ and\ \citenamefont {Sutton}}]{Dudarev1998}%
  \BibitemOpen
  \bibfield  {author} {\bibinfo {author} {\bibfnamefont {S.~L.}\ \bibnamefont
  {Dudarev}}, \bibinfo {author} {\bibfnamefont {G.~A.}\ \bibnamefont {Botton}},
  \bibinfo {author} {\bibfnamefont {S.~Y.}\ \bibnamefont {Savrasov}}, \bibinfo
  {author} {\bibfnamefont {C.~J.}\ \bibnamefont {Humphreys}}, \ and\ \bibinfo
  {author} {\bibfnamefont {A.~P.}\ \bibnamefont {Sutton}},\ }\href@noop {}
  {\bibfield  {journal} {\bibinfo  {journal} {Physical Review B}\ }\textbf
  {\bibinfo {volume} {57}} (\bibinfo {year} {1998})}\BibitemShut {NoStop}%
\bibitem [{\citenamefont {Dudko}\ \emph {et~al.}(1971)\citenamefont {Dudko},
  \citenamefont {Eremenko},\ and\ \citenamefont {Semenenko}}]{Dudko1971}%
  \BibitemOpen
  \bibfield  {author} {\bibinfo {author} {\bibfnamefont {K.~L.}\ \bibnamefont
  {Dudko}}, \bibinfo {author} {\bibfnamefont {V.~V.}\ \bibnamefont {Eremenko}},
  \ and\ \bibinfo {author} {\bibfnamefont {L.~M.}\ \bibnamefont {Semenenko}},\
  }\href {\doibase 10.1002/pssb.2220430203} {\bibfield  {journal} {\bibinfo
  {journal} {physica status solidi (b)}\ }\textbf {\bibinfo {volume} {43}},\
  \bibinfo {pages} {471} (\bibinfo {year} {1971})}\BibitemShut {NoStop}%
\bibitem [{\citenamefont {Tobia}\ \emph {et~al.}(2010)\citenamefont {Tobia},
  \citenamefont {Biasi}, \citenamefont {Granada}, \citenamefont {Troiani},
  \citenamefont {Zampieri}, \citenamefont {Winkler},\ and\ \citenamefont
  {Zysler}}]{Tobia2010}%
  \BibitemOpen
  \bibfield  {author} {\bibinfo {author} {\bibfnamefont {D.}~\bibnamefont
  {Tobia}}, \bibinfo {author} {\bibfnamefont {E.~D.}\ \bibnamefont {Biasi}},
  \bibinfo {author} {\bibfnamefont {M.}~\bibnamefont {Granada}}, \bibinfo
  {author} {\bibfnamefont {H.~E.}\ \bibnamefont {Troiani}}, \bibinfo {author}
  {\bibfnamefont {G.}~\bibnamefont {Zampieri}}, \bibinfo {author}
  {\bibfnamefont {E.}~\bibnamefont {Winkler}}, \ and\ \bibinfo {author}
  {\bibfnamefont {R.~D.}\ \bibnamefont {Zysler}}\ }(\bibinfo {year}
  {2010})\BibitemShut {NoStop}%
\bibitem [{\citenamefont {Lübbe}\ and\ \citenamefont
  {Moritz}(2009)}]{Lubbe2009}%
  \BibitemOpen
  \bibfield  {author} {\bibinfo {author} {\bibfnamefont {M.}~\bibnamefont
  {Lübbe}}\ and\ \bibinfo {author} {\bibfnamefont {W.}~\bibnamefont
  {Moritz}},\ }\href {\doibase 10.1088/0953-8984/21/13/134010} {\bibfield
  {journal} {\bibinfo  {journal} {Journal of Physics Condensed Matter}\
  }\textbf {\bibinfo {volume} {21}} (\bibinfo {year} {2009}),\
  10.1088/0953-8984/21/13/134010}\BibitemShut {NoStop}%
\bibitem [{\citenamefont {Bikondoa}\ \emph {et~al.}(2010)\citenamefont
  {Bikondoa}, \citenamefont {Moritz}, \citenamefont {Torrelles}, \citenamefont
  {Kim}, \citenamefont {Thornton},\ and\ \citenamefont
  {Lindsay}}]{Bikondoa2010}%
  \BibitemOpen
  \bibfield  {author} {\bibinfo {author} {\bibfnamefont {O.}~\bibnamefont
  {Bikondoa}}, \bibinfo {author} {\bibfnamefont {W.}~\bibnamefont {Moritz}},
  \bibinfo {author} {\bibfnamefont {X.}~\bibnamefont {Torrelles}}, \bibinfo
  {author} {\bibfnamefont {H.~J.}\ \bibnamefont {Kim}}, \bibinfo {author}
  {\bibfnamefont {G.}~\bibnamefont {Thornton}}, \ and\ \bibinfo {author}
  {\bibfnamefont {R.}~\bibnamefont {Lindsay}},\ }\href {\doibase
  10.1103/PhysRevB.81.205439} {\bibfield  {journal} {\bibinfo  {journal}
  {Physical Review B - Condensed Matter and Materials Physics}\ }\textbf
  {\bibinfo {volume} {81}} (\bibinfo {year} {2010}),\
  10.1103/PhysRevB.81.205439}\BibitemShut {NoStop}%
\bibitem [{\citenamefont {Kaspar}\ \emph {et~al.}(2013)\citenamefont {Kaspar},
  \citenamefont {Chamberlin},\ and\ \citenamefont {Chambers}}]{Kaspar2013}%
  \BibitemOpen
  \bibfield  {author} {\bibinfo {author} {\bibfnamefont {T.~C.}\ \bibnamefont
  {Kaspar}}, \bibinfo {author} {\bibfnamefont {S.~E.}\ \bibnamefont
  {Chamberlin}}, \ and\ \bibinfo {author} {\bibfnamefont {S.~A.}\ \bibnamefont
  {Chambers}},\ }\href {\doibase 10.1016/j.susc.2013.09.005} {\bibfield
  {journal} {\bibinfo  {journal} {Surface Science}\ }\textbf {\bibinfo {volume}
  {618}},\ \bibinfo {pages} {159} (\bibinfo {year} {2013})}\BibitemShut
  {NoStop}%
\end{thebibliography}%

\end{document}